\documentclass[aip,pop,amsmath,amssymb, reprint]{revtex4-1}
\usepackage{graphicx}
\usepackage{dcolumn,bm}
\usepackage[english]{babel}
\usepackage{sidecap}
\usepackage[T1]{fontenc}
\usepackage[colorlinks,linkcolor=blue,citecolor=blue,urlcolor=blue]{hyperref}

\hypersetup{pdftitle={Gyrokinetic and kinetic particle-in-cell simulations of guide-field reconnection. I. Macroscopic effects of the electron flows},pdfauthor={P. A. Mu\~noz}}

\begin{document}
\title[]{
	Gyrokinetic and  kinetic particle-in-cell simulations of guide-field reconnection. I. Macroscopic effects of the electron flows
}

\author{P. A. Mu\~noz}
\email{munozp@mps.mpg.de}
\affiliation{Max-Planck-Institut f\"ur Sonnensystemforschung, D-37077 G\"ottingen, Germany}

\author{D. Told}
\affiliation{Max-Planck-Institut f\"ur Plasmaphysik,  D-85748 Garching, Germany}
\affiliation{Department of Physics and Astronomy,  University of California,  Los Angeles,  California 90095, USA}

\author{P. Kilian}
\affiliation{Max-Planck-Institut f\"ur Sonnensystemforschung, D-37077 G\"ottingen, Germany}

\author{J. B\"uchner}
\affiliation{Max-Planck-Institut f\"ur Sonnensystemforschung, D-37077 G\"ottingen, Germany}

\author{F. Jenko}
\affiliation{Max-Planck-Institut f\"ur Plasmaphysik,  D-85748 Garching, Germany}
\affiliation{Department of Physics and Astronomy,  University of California,  Los Angeles,  California 90095, USA}

\date{\today}

\begin{abstract}

	In this work, we compare gyrokinetic (GK) and fully kinetic Particle-in-Cell (PIC) simulations of magnetic reconnection in the limit of strong guide field.
			In particular,  we analyze the limits of applicability of the GK plasma model  compared to a fully kinetic description of force free current sheets for finite guide fields ($b_g$).
	Here we report the first part of an extended comparison, focusing on the macroscopic effects of the electron flows.
	For a low beta plasma ($\beta_i=0.01$), it is shown that both plasma models develop magnetic reconnection with  similar features  in the secondary magnetic islands if a sufficiently high guide field ($b_g\gtrsim 30$) is imposed in the kinetic PIC simulations. Outside of these regions,  in the separatrices close to the X points,  the convergence between both plasma descriptions is less restrictive ($b_g\gtrsim 5$).
	Kinetic PIC simulations using guide fields $b_g \lesssim 30$ reveal secondary magnetic islands with a core magnetic field and less energetic flows inside of them in comparison to the GK or kinetic PIC runs with stronger guide fields.
	We find that these processes are mostly due to an initial shear flow absent in the GK initialization and negligible in the kinetic PIC high guide field regime,  in addition to fast outflows on the order of the ion thermal speed that violate the GK ordering.
	Since secondary magnetic islands appear after the reconnection peak time,  a kinetic PIC/GK comparison  is more accurate in the linear phase of magnetic reconnection.
	For a high beta plasma ($\beta_i=1.0$) where reconnection rates and fluctuations levels are reduced,  similar processes happen in the secondary magnetic islands in the fully kinetic  description, but requiring much lower guide fields ($b_g\lesssim 3$).\\

	\textit{Copyright (2015) American Institute of Physics. This article may be downloaded for personal use only. Any other use requires prior permission of the author and the American Institute of Physics.\\
The following article appeared in  P. A. Mu\~noz, D. Told, P. Kilian, J. B\"uchner and F. Jenko, Physics of Plasmas \textbf{22}, 082110 (2015),   and may be found at
\href{http://dx.doi.org/10.1063/1.4928381}{http://dx.doi.org/10.1063/1.4928381}
 }
\end{abstract}
\maketitle
\section{Introduction}\label{sec:intro}
Magnetic reconnection is a fundamental process in plasma physics that converts magnetic field energy efficiently into plasma heating, kinetic energy of bulk flows and accelerated particles. It plays a role in a wide range of different environments, from fusion devices,  planetary magnetospheres, and stellar coronae to extragalactic accretion disks.\cite{Buchner2007, Zweibel2009, Yamada2010}

Originally,  magnetic reconnection was modeled with antiparallel magnetic fields in a plane sustained by a current sheet (CS). But that configuration is not common in nature, since the presence of an  out-of-plane magnetic field is ubiquitous in both space and laboratory plasmas,  such as in the solar corona and fusion devices,  respectively. However,  the properties of reconnection in the presence of a guide field ($b_g$) are still nowadays much less understood.\cite{Treumann2013b} Thus,  the study of stability of these CS with shear magnetic fields is of paramount importance to understand the conversion of magnetic energy in these environments.

In order to capture the kinetic effects in magnetic reconnection, fully kinetic particle-in-cell (PIC) codes have become one of the preferred tools since some time ago,  with the increasing availability of computational resources.\cite{Buchner2003} These codes allow the simulation of magnetic reconnection in fully collisionless Vlasov plasmas. However, these fully kinetic PIC simulations (for brevity,  from now on these ones will be simply referred to as ``PIC simulations'') can be computationally very demanding due to stability reasons and the requirement of keeping numerical noise as low as possible.
One approach to solve this problem is using gyrokinetic (GK) theory,  an approximation to a Vlasov plasma (see Ref.~\onlinecite{Frieman1982} or Refs.~\onlinecite{Brizard2007,Howes2006,Schekochihin2009} for more recent reviews). In this plasma model,
the idea is to decouple the fast gyrophase dependence from the dynamics, considering the motion of charged rings,  thus reducing the phase space from six to five dimensions and removing dynamics on very small space-time scales. %
The gyrokinetic approach is suitable in strongly magnetized plasmas (equivalent to the limit of very large magnetic guide fields in magnetic reconnection). More precisely, the ion gyroradius $\rho_i$ has to be much smaller than the typical length scale $L_0$ of the background: $\epsilon=\rho_i/ L_0\ll 1$,  where $\epsilon$ is the GK ordering parameter. In contrast, it is important to note that the  typical length scale of variation of the perpendicular perturbations, $k_{\perp}^{-1}$, can be comparable to the ion gyroradius: $k_{\perp}\rho_i\sim 1$. The ordering $\epsilon\ll 1$ implies the restriction of the GK approach to phenomena with frequencies lower than the ion cyclotron one: $\omega/ \Omega_{ci}\sim\mathcal{O}(\epsilon)$,  where $\omega$ is a typical frequency in the system. The fluctuations in the distribution function and electromagnetic fields,  with respect to their equilibrium values,  are also of order $\mathcal{O}(\epsilon)$. And finally, these fluctuations are assumed to have much larger scales along the magnetic field direction than across it: $k_{\parallel}/k_{\perp}\sim\mathcal{O}(\epsilon)$,  where $k_{\parallel}$ and $k_{\perp}$ are the wavevectors along and across the magnetic field,  respectively. An important consequence of this assumption is that the perpendicular bulk speed $\vec{V}_{\perp}$ is  mostly dominated by drifts, which have to be smaller than the ion thermal speed $v_{th,i}$ according to $\epsilon=V_{\perp}/v_{th,i}\ll1$. This fact will be a source of differences between the GK and kinetic simulations that do not have such a restriction.

Simulations of magnetic reconnection with codes based in the GK theory have shown to drastically reduce the computation time, which allows the use of more realistic numerical parameters (e.g.:  mass ratio), and at the same time,  the possibility to capture many of the desired kinetic effects,  such as finite ion Larmor radius ones.\cite{Wan2005a, Rogers2007, Pueschel2011, Pueschel2014a} However,  one of the drawbacks of the GK simulations of magnetic reconnection is that the initial setup has to be force free $\vec{J}\times\vec{B}=0$. The often used Harris sheet\cite{Harris1962} initialization is not possible in the standard gyrokinetic theory,  since the particles that carry the current are different from the background,  and therefore the perturbed current is of second order in the ordering  GK parameter $\epsilon$. Nevertheless,  a force free initialization is a good approximation to the exact kinetic equilibrium in low beta plasmas with strong guide field.
Due to the previous reasons, it is of fundamental importance to establish the key properties of magnetic reconnection that can be properly modeled  with a gyrokinetic code, instead of running computationally expensive fully kinetic PIC simulations. In this context, the first direct comparison between PIC and GK simulations of magnetic reconnection in the large guide field limit was performed just recently in Ref.~\onlinecite{TenBarge2014}.
They found that  many reconnection related fluctuating quantities (such as the thermal/magnetic pressures) obtained for different PIC guide field runs have the same value after they are scaled to $b_g$. And this value is equal to the result given by the GK simulations after a proper choice of the $\epsilon$ parameter.  This is valid under the assumption of a constant total plasma $\beta$  (to be defined in Sec.~\ref{sec:setup}) among different PIC guide field runs.  The result is in agreement with the predictions of a two fluid analysis of Ref.~\onlinecite{Rogers2003}. In addition, Ref.~\onlinecite{TenBarge2014} showed that  the morphology of the out-of-plane current density exhibits a good convergence between the PIC simulations with sufficiently high guide field and the gyrokinetic runs. They also noticed that low $\beta$ plasmas require higher PIC guide fields to reach convergence towards the GK results, in comparison to high $\beta$ plasmas.
However,  the previous benchmark study in Ref.~\onlinecite{TenBarge2014} did not discuss some limitations in the GK approach,  that might make a comparison with a fully kinetic description of magnetic reconnection not reliable in some parameter regimes. In this context, our purpose is investigating the physical origin of the differences between these plasma models in the realistic regime of finite PIC guide fields (since PIC cannot use arbitrarily large values of guide field,  where results should match). Thus, we aim to establish the limits of applicability of the GK compared to PIC simulations of magnetic reconnection.
Some of the differences come from the special way in which the GK plasma model enforces the force balance in the perpendicular direction to the magnetic field to order $\epsilon$. Indeed, from the  perpendicular GK Amp\`ere's law, it can be proven\cite{Roach2005,Schekochihin2009,Abel2013}
\begin{align}\label{pressure_equilibrium_gk}
\nabla_{\perp} \cdot \delta  \overline{\overline{P_{\perp}}}  + \frac{B \nabla_{\perp} \delta B_{\parallel}}{2\mu_0} =0.
\end{align}
Here,  the quantities denoted as $\delta$ are of order $\epsilon$. $\delta  \overline{\overline{P_{\perp}}}$ is the (total) perturbed perpendicular pressure tensor and $\delta B_{\parallel}$ the perturbed parallel magnetic field. This expression is satisfied exactly to order $\epsilon$, being formally equivalent to the pressure equilibrium condition  as given by the two fluid model in the strong guide field limit (to be discussed in Eq.~\eqref{pressure_equilibrium}). The only difference is that $\delta  \overline{\overline{P_{\perp}}}$ can possibly include off-diagonal elements representing finite Larmor radius effects in the GK model, while in the two-fluid model the thermal pressure is only a scalar quantity. Note that even though the  magnetic pressure  is much larger than the  thermal pressure in this force free scenario, their gradients or fluctuations can still be comparable, and that is why such pressure equilibrium still exists even in the strong guide field limit as assumed in the GK approach. On the other hand, there is no such restriction in the PIC approach, but it  converges towards it as the guide field increases.

The physical consequence of the previous fact is that the agreement between both approaches breaks down mostly when secondary magnetic islands, with strongly unbalanced magnetic pressures, appear in the PIC simulations using a (relatively) low guide field. Although they also appear in the PIC high guide field regime or the GK simulations,  these secondary magnetic islands have different features as a result of an initial shear flow present in the PIC force free initialization.
These secondary magnetic islands are structures that appear as a result of the tearing mode, with a different (smaller) wavelength than the one imposed by the large scale initial perturbation. They start at electron length scales,  growing up to ion scales. In this sense, our definition does not exactly match with some previous studies,\cite{Chen2012,Zhou2012,Huang2014c} which limit secondary islands to those ones with electron length scales. Other works have stated that secondary islands have opposite out-of-plane current density to those of the primary islands.\cite{Huang2013} Again, the structures in our simulations do not match these features.

We expect the formation of these structures in our setup, since it is known that guide field reconnection produce a ``burstier'' reconnection,  forming more secondary magnetic islands than an antiparallel configuration.\cite{Drake2006} Many previous works have analyzed these structures with several levels of details,  in particular in the context of extended current layers,  since they can modulate temporally the reconnection rates (see Ref.~\onlinecite{Daughton2011c} and references therein).  For example, the authors of Ref.~\onlinecite{Karimabadi1999} reported a core magnetic field inside of secondary magnetic islands by means of hybrid simulations of Harris sheets with a weak magnetic guide field. It was attributed to a nonlinear amplification mechanism between the guide field and the out-of-plane magnetic field generated by Hall currents. Later,  \citet{Zhou2014} analyzed the same phenomenon  during the coalescence of secondary magnetic islands, with 2D PIC simulations for a similar setup. In our case, we will show that a similar consequence takes place but as a result of a shear flow initialization in the PIC low guide field regime.

Shear flows have been analyzed exhaustively for their importance in the development of magnetic reconnection. It is known that reconnection rates decrease with faster shear flows parallel to the reconnected magnetic field, according to the scaling law derived in Ref.~\onlinecite{Cassak2011a} under a Hall-MHD model without guide field (basically, due to less efficient outflows from the X point). A kinetic approach and PIC simulations have confirmed this prediction,  but  have also shown that tearing mode (associated with the formation of magnetic islands) is never completely stabilized for thin CS,  even with super-Alfv\'enic flows.\cite{Roytershteyn2008} A two-fluid analytical study\cite{Hosseinpour2013} of collisionless tearing mode driven by electron inertia showed that a shear flow,  under certain conditions,  can generate a symmetric out-of-plane magnetic field. Shear flows are also closely related with the Kelvin-Helmholtz instability. Our PIC runs in the low guide field regime do not exhibit outflows with sufficient speed to drive the kinetic version of Kelvin-Helmholtz, mostly because they scale as the Alfv\'en speed $V_A$ which is known to be a threshold for this instability.\cite{Wang1992}
It is also important to mention the work by \citet{Fermo2012} and \citet{Liu2014}, which found,  by means of 2D PIC simulations,  secondary magnetic islands generated by the Kelvin-Helmholtz instability in the reconnection outflow (similar to the finding by \citet{Loureiro2013a},  but in the framework of reduced MHD). On the other hand, the structures observed in our case appear only close to the main X point. However, the secondary magnetic islands in our PIC/GK simulations share some of the features seen in those previous works. That is why we are going to relate their appearance with the core magnetic field and shear flow,  depending on the parameter regimes of our PIC/GK runs.

The remainder of this paper is organized as follows. In Sec.~\ref{sec:setup}, we describe the simulation setup used by both GK and PIC codes, as well as the parameter range to be studied. In Sec.~\ref{sec:linearscaling} we identified the deviations from the linear scaling on the guide field of some magnetic reconnection quantities,  extending the previous comparison work by \citet{TenBarge2014}
These differences  between PIC and GK simulations,  appearing for some parameter regimes and specific spatial and temporal locations,  are manifested in both magnetic and thermal pressure fluctuations.
In this paper we focus on the first ones, while the deviations with respect to the linear scaling in the thermal pressure fluctuations will be deferred to a follow-up paper.
The remarkably high values of magnetic fluctuations in the PIC low guide field regime are due to the generation of a core magnetic field in the secondary magnetic islands. This process is discussed in  Sec.~\ref{sec:corefield_pressureequilibrium} in the context of two fluid theory and violations of the pressure equilibrium condition. Then, in Sec.~\ref{sec:bzgeneration}, we analyze the physical mechanism of this core magnetic field as a result of a shear flow due to the force free initialization in the PIC low guide field runs. We briefly discuss the effects of a high $\beta$ plasma on all these processes in Sec.~\ref{sec:finite_plasmabeta}. Finally, we summarize our findings in the conclusion, Sec.~\ref{sec:conclusions}.
\section{Simulation setup}\label{sec:setup}
The results of the 2.5D simulations (no variations allowed in the out-of-plane direction $\hat{z}$) to be shown were obtained with the explicit fully kinetic PIC code ACRONYM\cite{Kilian2012} and the gyrokinetic GENE code.\cite{Jenko2000} For both PIC and gyrokinetic simulations, we used a very similar setup and parameter set to the ones used in Ref.~\onlinecite{TenBarge2014}. We will repeat here the more important parameters to ensure a future accurate reproducibility of our work. The initial equilibrium is a force free double CS, with halfwidth $L$ and magnetic field $\vec{B}(x)=B_y(x)\hat{y} + B_z(x)\hat{z}$ given by:
{\small
\begin{align}
	B_y &= B_{\infty y}\left[ \tanh\left(\frac{x-L_x/4}{L}\right) -  \tanh\left(\frac{x-3L_x/4}{L}\right) -1 \right],  \label{by_initial}\\
	B_z &= B_{\infty y} \left[b_g^2 + \cosh^{-2}\left(\frac{x-L_x/4}{L}\right)
+ \cosh^{-2}\left(\frac{x-3L_x/4}{L}\right)\right]^{1/2},\label{bz_initial}
\end{align}
}
with relative guide field strength $b_g=B_g/B_{\infty y}$, where $B_{\infty y}$ is the (asymptotic) reconnected magnetic field.  This corresponds to a total magnetic field of constant magnitude $B_T=B_{\infty y}\sqrt{1+b_g^2}$. A change in $b_g$ is obtained by modifying $B_{\infty y}$ but always keeping $B_T$ constant (note that  Ref.~\onlinecite{TenBarge2014} keeps $B_g$ constant, but $B_T\sim B_g$ if $b_g\gg1$ and so the differences are not significant in this regime). $L_x$ and $L_y$ are the simulation box sizes in the reconnection plane $x$-$y$,  respectively. Ions are initially stationary,  while the current is carried by drifting electrons with velocity $\vec{U}_e$ according to $\vec{J}=-en_e\vec{U}_e$,  satisfying the Amp\`ere's law $\nabla\times \vec{B}=\mu_0 \vec{J}$. $n_e=n_i=n_0$ are the initial constant electron (``e'') and proton (``i'') number densities,  respectively. In order to accelerate the reconnection onset, we use a perturbation in $B_x$ and $B_y$ that generates an X/O point at the center of the left/right CS and that can be derived from the vector potential $\delta A_z$
\begin{align}\label{perturbation}
	\delta A_z = \delta P \frac{L_y}{2\pi}\sin\left(\frac{2\pi \left(y+L_y/4\right)}{L_y}\right)\sin^2\left(\frac{2\pi x}{L_x}\right),
\end{align}
where $\delta P=0.01$ is the amplitude of the perturbation. The parameters used in this study are divided in two sets: ``low'' and ``high'' beta cases,  being summarized in Table~\ref{tab:table}. All the corresponding quantities are calculated with respect to $B_T$,  i.e.: the ion plasma beta $\beta_i = 2\mu_0 n_0 k_B T_i/B_T^2$, the electron/ion cyclotron frequency  $\Omega_{c\{e/ i\}}=eB_T/m_{\{e/i\}}$ and the electron/ion Larmor radius $\rho_{\{e/i\}}=v_{th,{\{e/i\}}}/\Omega_{c\{e/ i\}}$,  with the electron/ion thermal speed $v_{th,{\{e/i\}}}=\sqrt{2k_BT_{\{e/ i\}}/m_{\{e/ i\}}}$. We also have equal temperatures for both species $T_i=T_e$, implying $\sqrt{\beta_i}=(\omega_{pe}/\Omega_{ce})(v_{th, e}/c)$. All the details concerning normalization and the appropriate correspondence between GK and PIC results can be found in the Appendix~\ref{sec:app_normalization}.
\begin{table}
	\begin{ruledtabular}
		\begin{tabular}{cccccc}
			Case      & $\beta_i$ & $\omega_{pe}/\Omega_{ce}$ & $v_{th,e}/c$ & $L/\rho_i$ & $L_x(=L_y)$   \\ \hline
			Low beta  & 0.01      & 0.8                       & 0.125        & 2          & 40$\pi\rho_i$ \\
			High beta & 1.0       & 4.0                       & 0.25         & 1          & 20$\pi\rho_i$ \\
		\end{tabular}
	\end{ruledtabular}
	\caption{\label{tab:table} Physical parameters for the two sets of runs.}
\end{table}

Now, let us specify some numerical parameters. For all cases we use a reduced mass ratio $m_i/m_e=25$. Both codes use double periodic boundary conditions ($x$ and $y$ directions). For the PIC runs,  we use a grid size $\Delta x$ such as $N_x=N_y=1024$ cells in the low beta case (with $\rho_e/\Delta x=1.69$),  while for the high beta case is $N_x=N_y=1280$ cells (with $\rho_e/\Delta x=4.07$). The time step is chosen to be $\Delta t\omega_{pe}=0.03/0.12$ in the low/high beta case,  respectively,  to fulfill the Courant-Friedrichs-Lewy (CFL) condition (for light waves) with $(c\Delta t)/\Delta x=0.5<1$. Thousand particles per cell are used in both cases for each species. A second order interpolation scheme, also known as TSC shape function (Triangular Shaped Cloud) was used to reduce numerical noise without having to increase significantly the macroparticle number. Finally, no current smoothing was used. For the GK runs,  the spatial grid is $N_x=N_y=1024$ for both cases,  while the parallel/perpendicular velocity grid is chosen to be $L_v=3v_{th, i}$,  $L_{\mu} = 9k_BT_i/B_T$,  with 32$\times$20 points in the space ($v_{\parallel},  \mu$). $\mu$ is the (adiabatic invariant) magnetic moment. In the GK simulations,  the initial noise level and spectrum were chosen to match with the corresponding one in the PIC runs. Then,  this noise acts as an additional perturbation on top of the one described in Eq.~\eqref{perturbation}.

Finally, some comments about the computational performance of the runs with our codes under the different parameter regimes of this study can be found in the Appendix~\ref{sec:app_times}.

\section{Global evolution and deviations from linear scaling}\label{sec:linearscaling}
In this section, we first show the global evolution of the  GK/PIC simulations comparing with the results obtained by \citet{TenBarge2014} Then,  we explain and quantify the linear scaling on the guide field  as predicted by the two fluid model of Ref.~\onlinecite{Rogers2003}, in order to identify the deviations from it and
the key open problems that will be addressed in this and an upcoming paper.
\subsection{Global evolution}
\begin{SCfigure*}[\sidecaptionrelwidth][!ht]
\centering
	\includegraphics[width=1.4\linewidth]{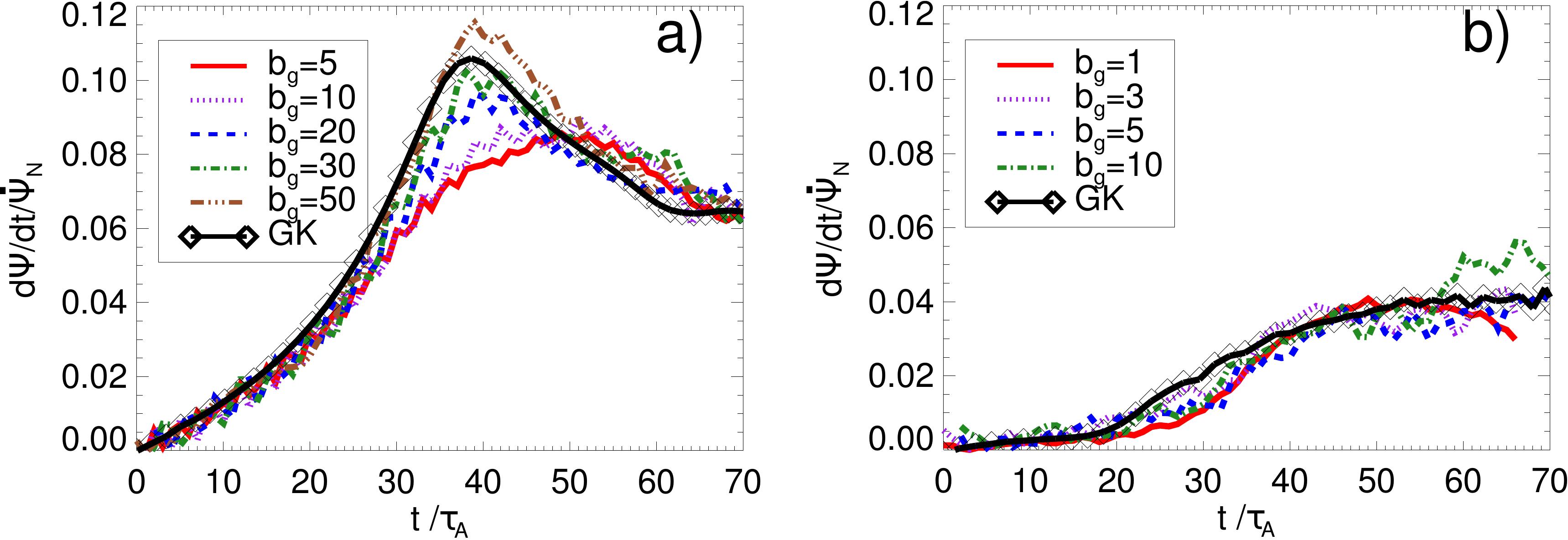}
	\caption{Comparison of reconnection rates $d\psi/dt$ for the left CS among different PIC guide field and GK runs. This quantity is calculated as the difference in the out-of-plane vector potential $A_z$ between the X and O points. a) Low beta $\beta_i=0.01$ case. b) High beta $\beta_i=1.0$ case.}\label{fig:recrates}
\end{SCfigure*}

With our independent set of codes, the PIC code ACRONYM and the GK code GENE, we obtained similar values for the reconnection rates (see Fig.~\ref{fig:recrates}) as the ones reported by the previous comparison work, Ref.~\onlinecite{TenBarge2014}. First, the normalized reconnection rates have the same value for each high/low $\beta$ case for different guide fields (although they decrease with $b_g$ when measured without normalization, consequence of keeping constant the total plasma $\beta$). The constancy of normalized reconnection rates in the strong guide field limit agrees with some previous studies, even when the total plasma $\beta$ changes\cite{Liu2014} (and correspondingly it was expected to have different reconnection regimes). Note that is not valid for Harris CS in the very low guide field regime, where it is known that the normalized reconnection rates are reduced with increasing $b_g$ (see, e.g., Refs.~\onlinecite{Pritchett2001,Pritchett2004,Ricci2004}), as a result of the enhanced incompressibility of the plasma.

Second, we also found smaller reconnection rates in the high beta case compared to the low beta one, in agreement with previous studies of gyrokinetic simulations of magnetic reconnection.\cite{Numata2014,Numata2014a} However, we can notice in Fig.~\ref{fig:recrates} that the reconnection values for both low and high plasma $\beta$ cases are still a substantial fraction of $(d\Psi/dt)/\dot{\Psi}_N\sim0.1$. Therefore, fast magnetic reconnection takes place in both cases, even though the low beta runs are in a regime where there should be not dispersive waves that may explain these rates, according to the two fluid analysis in Ref.~\onlinecite{Rogers2001} (see also Ref.~\onlinecite{Ricci2004}). This contradiction was already addressed in Ref.~\onlinecite{TenBarge2014} and also other recent works.\cite{Liu2014, Stanier2015, Cassak2015} Although there is no a full definitive answer to this controversy yet, all these evidences seem to indicate that fast dispersive waves, such as whistler and kinetic Alfv\'en waves, seem not to play the essential role to explain fast magnetic reconnection in collisionless plasmas. Since the purpose of this paper is different, we are not going to discuss further that issue in this work.

It is worth to mention that some details in the evolution of the system are not exactly the same as in Ref.~\onlinecite{TenBarge2014}. For instance, in the low beta case (Fig.~\ref{fig:recrates}(a)), although the peak in the reconnection rate is reached more or less at the same time, we observed that the CS is more prone to the development of secondary islands than reported there.  They start to appear just after the reconnection peak ($t\gtrsim 40\tau_A$) instead of $t\gtrsim 75\tau_A$.
Probably, this is due to the different initial noise level in the PIC runs, since it is random and dependent on the specifics of the algorithms implementation. Therefore,  the locations and timing of the secondary magnetic islands are expected to be different between the results given by the VPIC (as used in Ref.~\onlinecite{TenBarge2014}) and ACRONYM codes (and also with the GK code GENE).

All the results and plots to be shown from now on are based in the low beta case. The slightly different conclusions for the high beta case will be briefly analyzed only in  Sec.~\ref{sec:finite_plasmabeta}.

\subsection{Parity/symmetry of magnetic reconnection quantities and linear scaling}\label{sec:symmetry}

As indicated in Ref.~\onlinecite{TenBarge2014},  based in the two fluid analysis of Ref.~\onlinecite{Rogers2003}, the thermal pressure and magnetic field fluctuations should display antisymmetric (odd-parity) structures in the separatrices in the strong guide field regime $b_g\gg1$, assuming small enough fluctuations and asymptotic plasma beta $\beta_y=2\mu_0 n_0 k_B T_i/B_{\infty y}^2\gtrsim 1$ (although a later study\cite{Hosseinpour2013} showed that this assumption is not necessary for the appearance of an odd-parity magnetic field structure). In addition, both quantities should scale inversely with the guide field $b_g$ in the following way
(see Eqs. (17) and (19) of Ref.~\onlinecite{Rogers2003}):
\begin{align}\label{delta_thermalpressure}
	\frac{\delta P_{th}}{P_{th, 0}}& \sim \frac{d_i}{l_x}\frac{1}{b_g}, \\
	\frac{\delta B_z}{B_{g}} &\sim -\frac{\delta P_{th}}{B_{g}^2/\mu_0}\sim\frac{d_i}{l_x} \frac{\beta_i}{b_g}. \label{delta_magneticpressure}
\end{align}
 where $l_x$ is the typical length scale of variation of these quantities across the CS,  away enough from the X points. The fluctuations $\delta$ are calculated by subtracting the (initial) equilibrium quantities $B_z(t=0)\approx B_g$ (for $b_g\gg1$) and $P_{th}(t=0):=n_ek_B T_{e, \perp}+n_ik_B T_{i, \perp}$ (for brevity,  we omit the subscript $\perp$ in the thermal pressure).  Note that in deriving the previous expressions, the pressure equilibrium condition in the limit of strong guide field has been used\cite{Rogers2003} (also obtained in the framework of reduced MHD), formally equivalent (by assuming a scalar pressure) to the perpendicular force balance of the more general GK equations as given in Eq.~\eqref{pressure_equilibrium_gk} (satisfied exactly to order $\epsilon$). This can be written in the following way by combining the previous expressions for $\delta P_{th}$ and $\delta B_z$:
\begin{equation}\label{pressure_equilibrium}
	\frac{\delta B_z}{B_g}=-\beta_i\frac{\delta P_{th}}{P_{th, 0}}.
\end{equation}

\begin{figure*}[!ht]
	\centering
	\includegraphics[width=\linewidth]{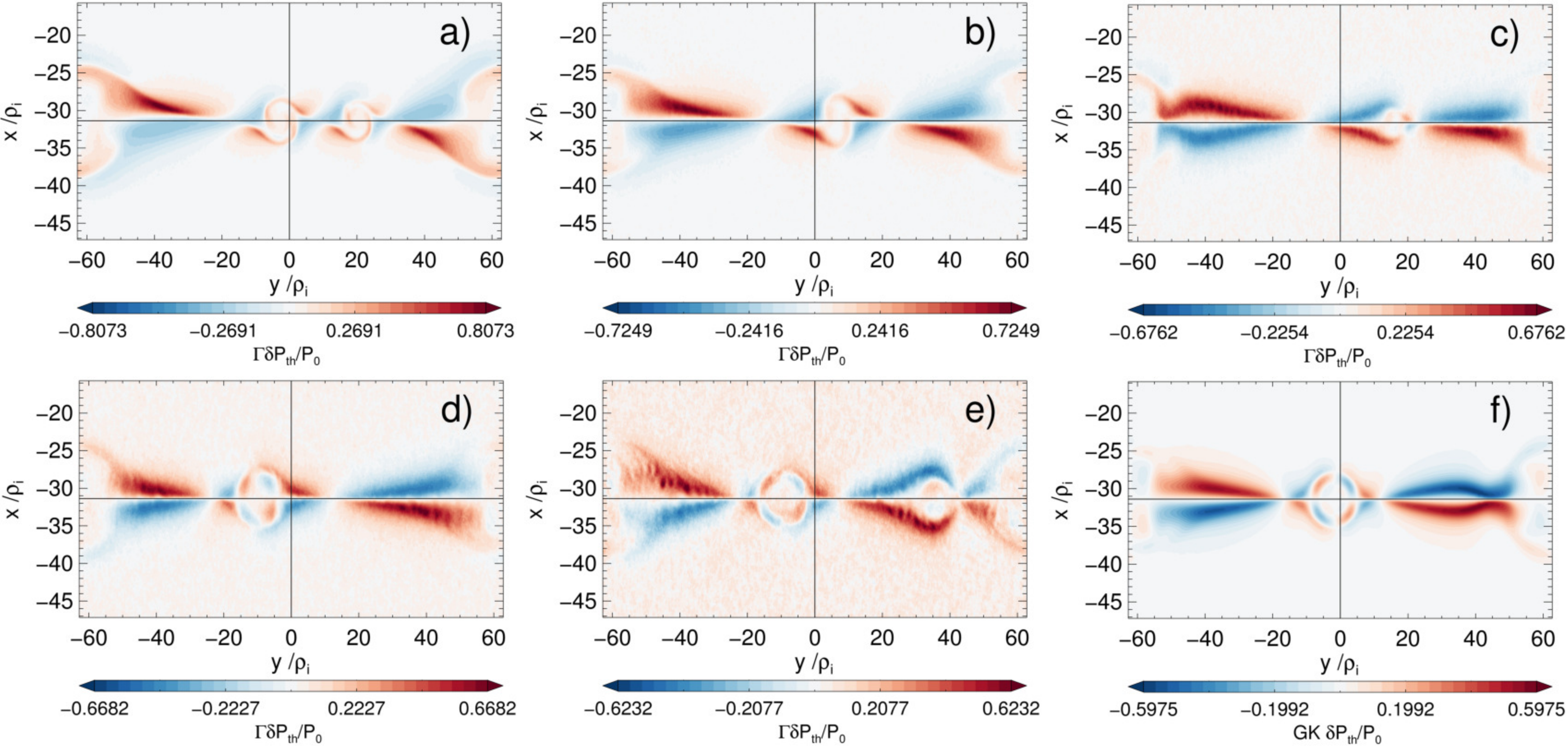}
	\caption{Contour plots of the scaled fluctuations in the (perpendicular) thermal pressure $\Gamma\delta P_{th}/P_{th, 0}$ for different PIC guide fields and GK runs,  at a time $t/\tau_A=50$. $\Gamma$ is for $b_{g,ref}=10$ in Eq.~\eqref{scaling_factor}. (The same is valid for all the remaining figures of this paper in the low beta case.) a) PIC $b_g=5$, b) PIC $b_g=10$, c) PIC $b_g=20$, d) PIC $b_g=30$, e) PIC $b_g=50$, f) GK. \label{fig:pth}}
\end{figure*}

\begin{figure*}[!ht]
	\centering
	\includegraphics[width=\linewidth]{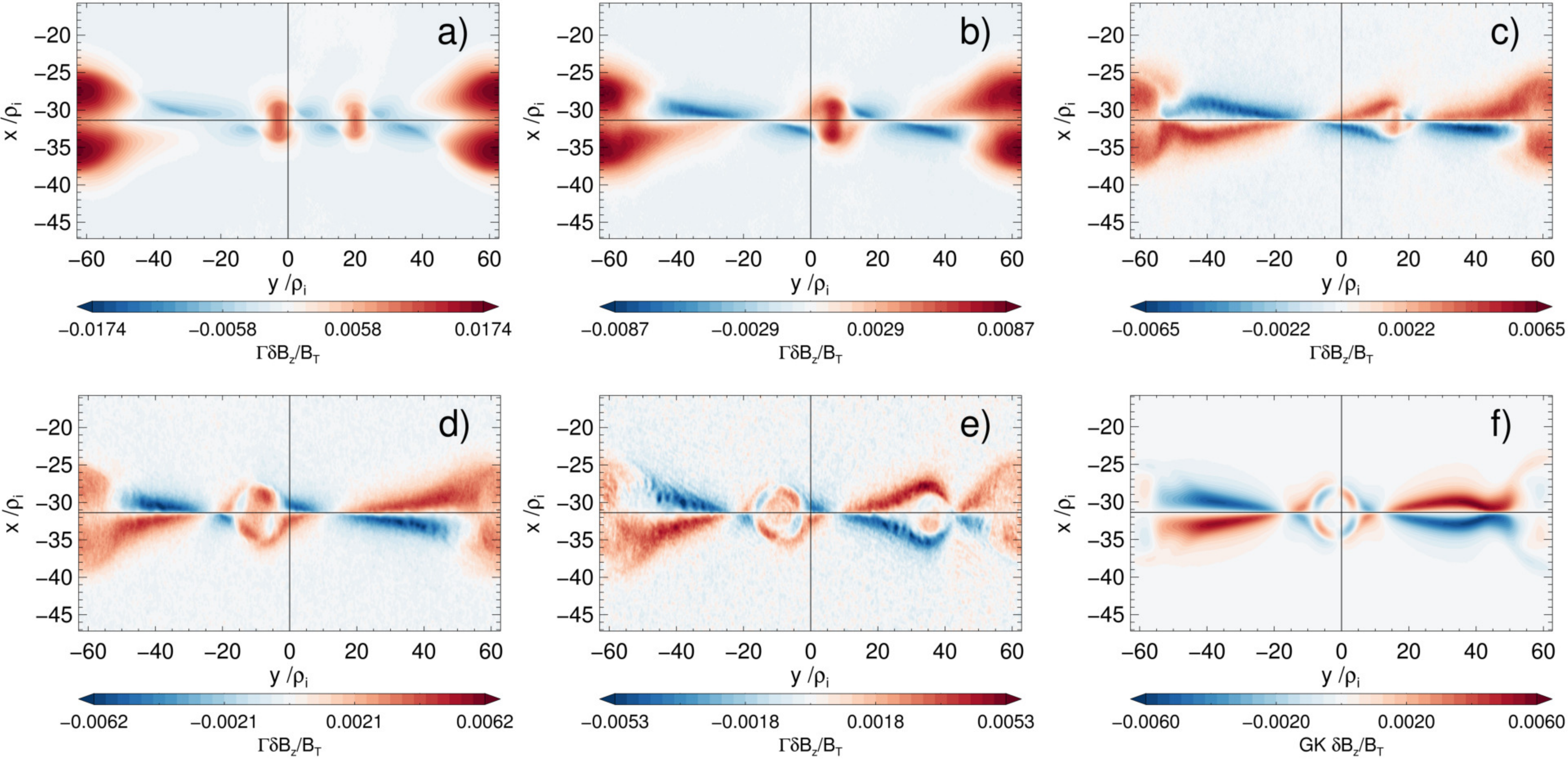}
	\caption{Same as Fig.~\ref{fig:pth} but for the scaled fluctuations in the out-of-plane magnetic field $\Gamma\delta B_z/B_T$. a) PIC $b_g=5$, b) PIC $b_g=10$, c) PIC $b_g=20$, d) PIC $b_g=30$, e) PIC $b_g=50$, f) GK.\label{fig:bz_contours}}
\end{figure*}

Note that Eqs.~\eqref{delta_thermalpressure} and \eqref{delta_magneticpressure} predict that the fluctuations $\delta P_{th}$ and $\delta B_z$ are proportional to $1/b_g$  providing $b_g\gg1$ and  $l_x/d_i$ constant for different guide fields, which is valid recurring to the estimate of order of magnitude $l_x/d_i \sim \sqrt{\beta_i}$ given in Ref.~\onlinecite{Rogers2003}.  The last assumption requires additional evidence in order to be applied to our simulations, and that is why we proved this claim via the method explained in detail in the Appendix~\ref{sec:app_transverse_distance}.
Using the suitable normalizations explained in Appendix~\ref{sec:app_normalization}, we show the estimates for $\delta B_z$ and $\delta P_{th}$ in Fig.~\ref{fig:pth} and Fig.~\ref{fig:bz_contours}, respectively, for a time shortly after the reconnection peak. These and all the other figures from the PIC runs shown in this paper, unless stated otherwise, have been averaged over $t=0.5\tau_A$ to reduce the effects of the numerical noise. In addition, the color scheme is scaled between $\pm$ the mean plus 3.5 standard deviations of the plotted quantity, a representative maximum as explained in the discussion of Fig.~\ref{fig:ff-linearscaling}.
From Figs.~\ref{fig:pth} and ~\ref{fig:bz_contours} we confirm the result already found in Ref.~\onlinecite{TenBarge2014} and predicted by the previously sketched two fluid model:\cite{Rogers2003} the convergence of the PIC results towards the GK ones in the limit of strong guide field, in both odd symmetry and scaling with the guide field. As discussed by these authors, this is valid only in the region close to the separatrices. However, we can immediately notice a fact not discussed in the previous comparison work: not only the symmetry between both separatrices is broken in the low guide field regime, but also the appearance of an additional magnetic field in the secondary magnetic islands not visible in GK or in the limit of strong PIC guide field. This is the main difference and new result to be the focus of the remainder of this work. For brevity, other differences between PIC/GK will be addressed in a follow-up paper.

\section{Core magnetic field and pressure equilibrium condition}\label{sec:corefield_pressureequilibrium}
The core magnetic field in the secondary magnetic islands,  as well as in the $y$ boundaries,  can be understood from several points of view. In this section, we will describe it in the framework of deviations from the two fluid model sketched in Sec.~\ref{sec:symmetry}, based on the pressure equilibrium condition.
\subsection{Pressure equilibrium condition in the strong guide field limit in GK/PIC}
\begin{figure*}[!ht]
	\centering
	\includegraphics[width=\linewidth]{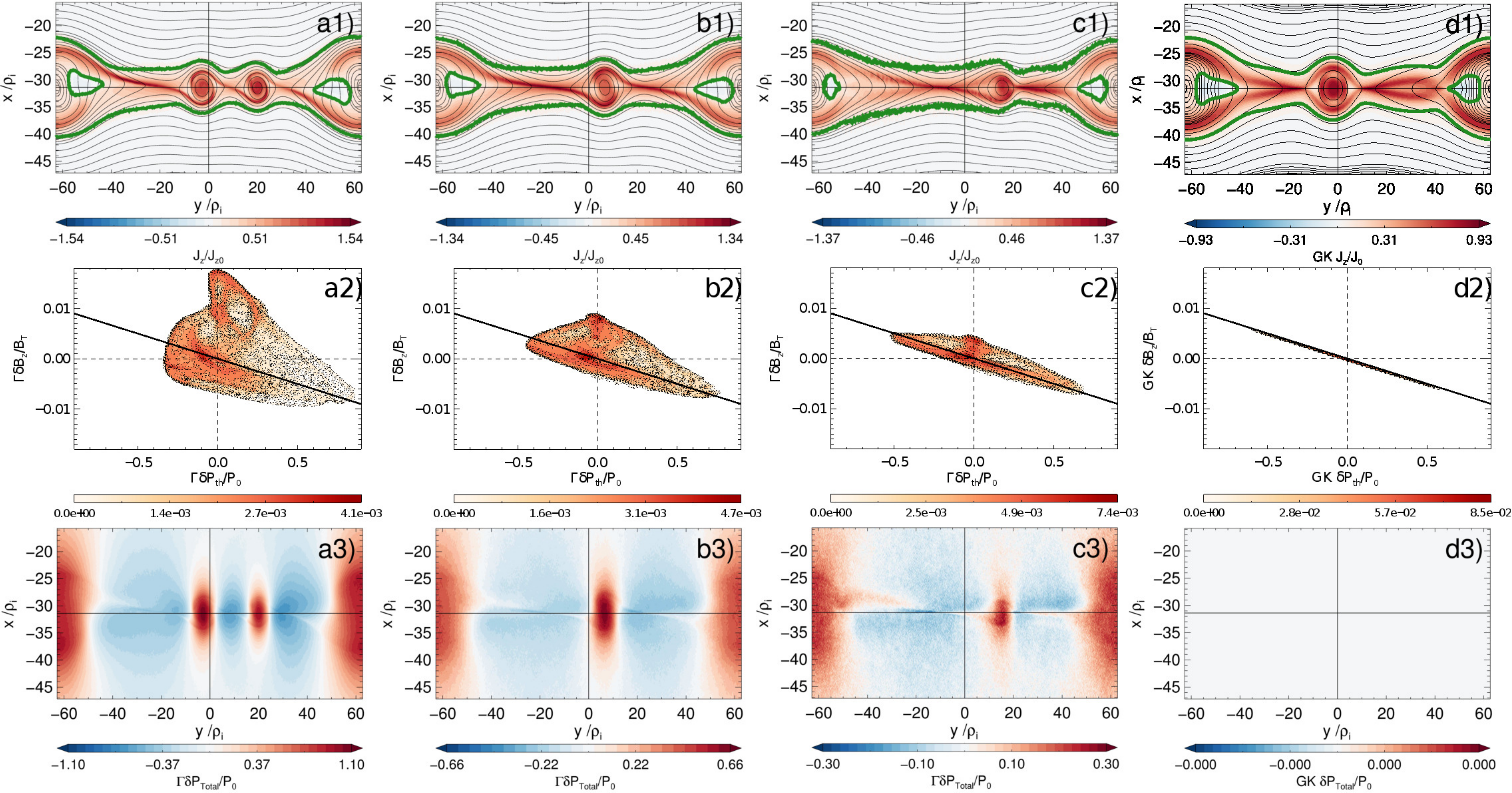}
	\caption{Quantities showing the method and locations of the deviations in the pressure equilibrium condition for different PIC guide fields and GK runs,  at a time $t/\tau_A=50$. Guide field increases to the right:  ax) PIC $b_g=5$, bx) PIC $b_g=10$, cx) PIC $b_g=20$, dx) GK, where the $x$ row is:\\
$x=1$ Top row: Contour plots of the out-of-plane current density $J_z/J_z(t=0)$. Magnetic field lines are shown in black contour lines.
The region inside of the green contour satisfies $J_z/J_z(t=0)>0.1$.\\
$x=2$ Middle row : 2D (frequency) histograms with the correlation between the magnetic and thermal fluctuations $\Gamma\delta B_z/B_T$ and $\Gamma\delta P_{th}/P_{th, 0}$.
The diagonal black straight line is
 the pressure equilibrium condition
Eq.~\eqref{pressure_equilibrium}. \\
$x=3$ Bottom row:  Contour plots for the scaled total pressure $\Gamma\delta P_{\rm total}/P_{th, 0}$. Note that $\delta P_{\rm total}=0$ to machine precision in GK (d3). \label{fig:ffbgs-pressurescorrelation}}
\end{figure*}

The core magnetic field in the secondary magnetic islands and $y$ boundaries  is unbalanced, in the sense that the thermal pressure does not decrease at these locations (compare Figs.~\ref{fig:pth} with \ref{fig:bz_contours}). This is a violation of the pressure equilibrium condition in the strong guide field limit as given in Eq.~\eqref{pressure_equilibrium_gk} or Eq.~\eqref{pressure_equilibrium}. Since it is satisfied exactly to order $\epsilon$ in GK, it is built-in in the equations that the GENE code solves, while PIC codes allow large deviations from it for finite guide fields, since they solve the full collisionless Vlasov-Maxwell system of equations. Note, in particular, that Eq.~\eqref{pressure_equilibrium} represents  a  straight line with slope $-\beta_i$ in the plane $\delta B_z$ - $\delta P_{th}$. Therefore, a convenient way to display this relation and deviations is by means of
a 2D (frequency) histogram of these fluctuating quantities, with results shown
in Fig.~\ref{fig:ffbgs-pressurescorrelation}(middle row).
These plots were generated by selecting the interesting region close enough to the center of the CS with  $J_z$ above $10\%$ of its initial value, indicated in  Fig.~\ref{fig:ffbgs-pressurescorrelation}(top row).
In order to show the locations of deviations in the pressure equilibrium condition,   we also plot in Fig.~\ref{fig:ffbgs-pressurescorrelation}(bottom row) the corresponding fluctuations in the total pressure
\begin{equation}\label{pressure_equilibrium2}
	\frac{\delta P_{\rm total}}{P_{th, 0}} = \frac{P_{th}+P_{mag}}{P_{th, 0}}-1-\frac{1}{2\beta_i},
\end{equation}
with $P_{mag}=B^2/(2\mu_0)$.
 Fig.~\ref{fig:ffbgs-pressurescorrelation}(middle row) shows
that  most of the locations in the CS are along the line $-\beta_i$ for the case PIC $b_g=20$, with an increasing spread in the lower $b_g$ regime. The corresponding GK results follow very accurately that line. Note a distinctive, pressure unbalanced, ``bump'' in the region $\delta P_{th}\sim 0$ with $\delta B_z\gtrsim0$, being very noticeably for the case PIC $b_g=5$.
Fig.~\ref{fig:ffbgs-pressurescorrelation}(bottom row) shows
 that these regions are mostly in the secondary magnetic islands and in the $y$ boundaries. This excess of total pressure generates a net force towards the exterior of the magnetic islands,  leading to an expansion in reconnection time scales ($\sim\tau_A$). $\delta P_{\rm total}$ increases going to the lower guide field regime,  even though this quantity has been linearly scaled to $b_g$.
That  is to be expected,  since the high fluctuation level in these cases breaks down the small $\epsilon$  assumption in which Eqs.~\eqref{delta_thermalpressure},\eqref{delta_magneticpressure} or \eqref{pressure_equilibrium} (and also Eq.~\eqref{pressure_equilibrium_gk}) are based.
Note that the histograms for low $b_g$ show a strongly asymmetric distribution in comparison with GK or  PIC high $b_g$ regime (see Fig.~\ref{fig:ffbgs-pressurescorrelation}(middle row)):
there are more points located in the right bottom quadrant ($\delta P_{th}>0$ and $\delta B_z<0$) than in the left upper quadrant ($\delta P_{th}<0$ and $\delta B_z>0$).
 This is an indication of (and an efficient way to quantify) the asymmetry in the separatrices for the PIC low guide field regime (see the respective contour plots Figs.~\ref{fig:pth} and \ref{fig:bz_contours}).
We will explain the reasons in Sec.~\ref{sec:shear_flow}.

We can summarize these findings by saying that,  although the magnetic field fluctuations $\delta B_z$ predicted by the GK simulations  can be comparable to the PIC ones in the low guide field regime ($b_g=5$ and $10$) close to the separatrices,  this is not true inside of the secondary magnetic islands or the periodic $y$ boundaries, due to the deviation in the pressure equilibrium Eq.~\eqref{pressure_equilibrium_gk}.
\subsection{Time evolution of magnetic and thermal fluctuations}

\begin{SCfigure*}[\sidecaptionrelwidth][!ht]
	\centering
	\includegraphics[width=1.4\linewidth]{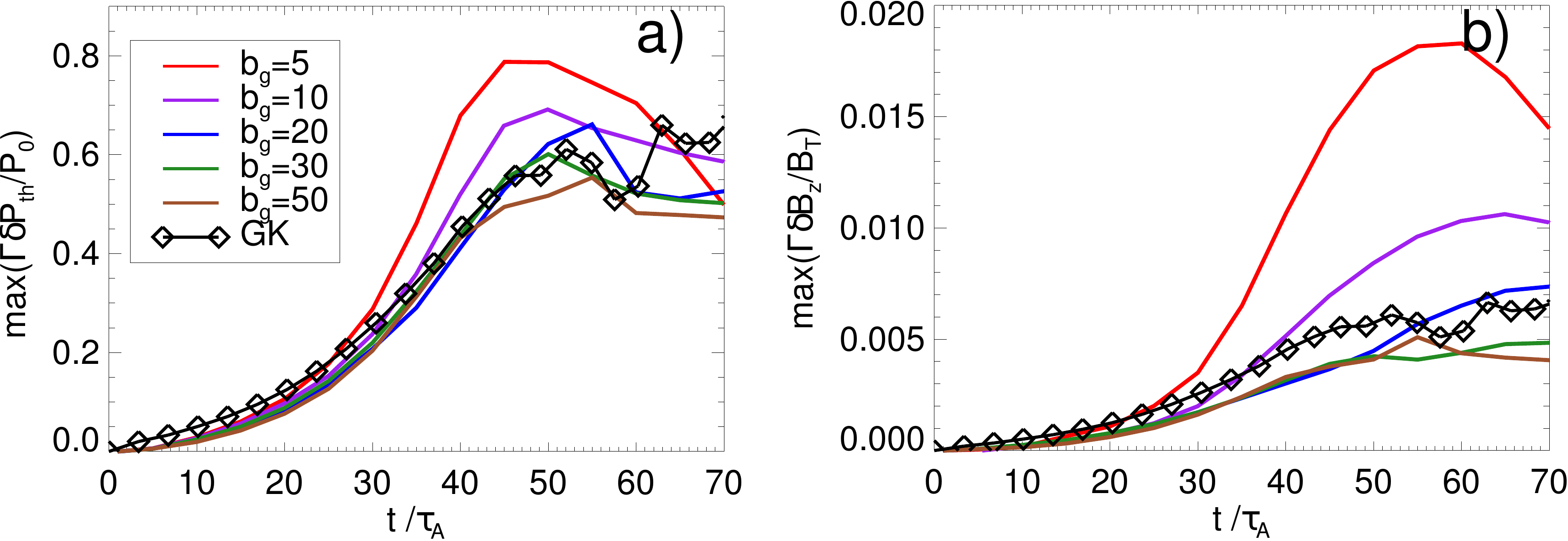}
	\caption{Time history of the ``maximum'' value of (a): scaled thermal pressure $\Gamma\delta P_{th}/P_{th, 0}$ and (b) magnetic fluctuations $\Gamma\delta B_z/B_T$ for different PIC guide fields and GK runs. See text for details about the calculation method.
\label{fig:ff-linearscaling}}
\end{SCfigure*}

The purpose of this subsection is to analyze the dynamical evolution of the thermal and magnetic fluctuations, complementing the work by \citet{TenBarge2014}. This allow us to determine the physical mechanisms producing the deviations in the pressure equilibrium condition, as well as the asymmetric separatrices in the PIC low guide field regime. We quantify this by means of tracking the
``maximum'' of $\delta P_{th}$ and $\delta B_z$ in the entire region  shown in Figs.~\ref{fig:pth} and  \ref{fig:bz_contours}. For the PIC simulations,  in order to decrease the effects of the numerical noise,  this  ``maximum'' is chosen as equal to the mean plus 3.5 standard deviations. The absolute maximum is not a good choice since  is very prone to outlier values.
In addition,  the initial value of the respective fluctuating quantity is subtracted (note that it is not only the initial value  $P_{th,0}$ and $B_{z,0}$), because: (1) a zero offset is required by GK since the fluctuating quantities are initially zero and (2) it mostly measures the numerical PIC noise,  being enhanced for higher $b_g$. The results are shown in Fig.~\ref{fig:ff-linearscaling}.

First of all, Fig.~\ref{fig:ff-linearscaling}(a) shows that the maximum of $\delta P_{th}$ is reached around $t\sim50\tau_A$ for all the cases, after the reconnection peak time $t\sim40\tau_A$ when also the secondary magnetic islands start to form.  $\delta B_z$ reaches maximum values even later. This in an indication of an additional mechanism generating $\delta B_z$,  different from the one produced by the Hall currents due to the reconnection process itself (close to the X point,  in the separatrices), and
deeply in the non-linear phase.
This is also the justification for showing the contour plots at a time $t=50\tau_A$ in most of the figures of this paper.

By means of the results shown in Figs.~\ref{fig:ff-linearscaling}(a) and ~\ref{fig:ff-linearscaling}(b), we also confirm (complementing the work of Ref.~\onlinecite{TenBarge2014}) the convergence of the time evolution of both PIC  $\Gamma \delta P_{th}$ and $\Gamma \delta B_{z}$  toward the GK curves in the strong guide field limit $b_g\gtrsim20$.
 As can be expected,  the GK values follow a similar trend as the strongest PIC guide field ($b_g=50$).
However, there are important deviations in some curves, noticeable earlier for smaller $b_g$. For example, the PIC run $b_g=5$ shows deviations from the GK curve of $\Gamma \delta B_{z}$  already in $t\gtrsim 20\tau_A$, while $b_g=30$ only after $t\gtrsim 45\tau_A$. Note that the deviations in $\Gamma \delta P_{th}$ are smaller than those of  $\Gamma \delta B_{z}$.
The (interesting) physical reasons of the differences for $\Gamma \delta P_{th}$
will be addressed in a follow-up paper.
Here we explain and analyze with more detail the differences in  $\Gamma\delta B_{z}$  shown in Fig.~\ref{fig:ff-linearscaling}(b).

\begin{figure*}[!ht]
	\centering
	\includegraphics[width=\linewidth]{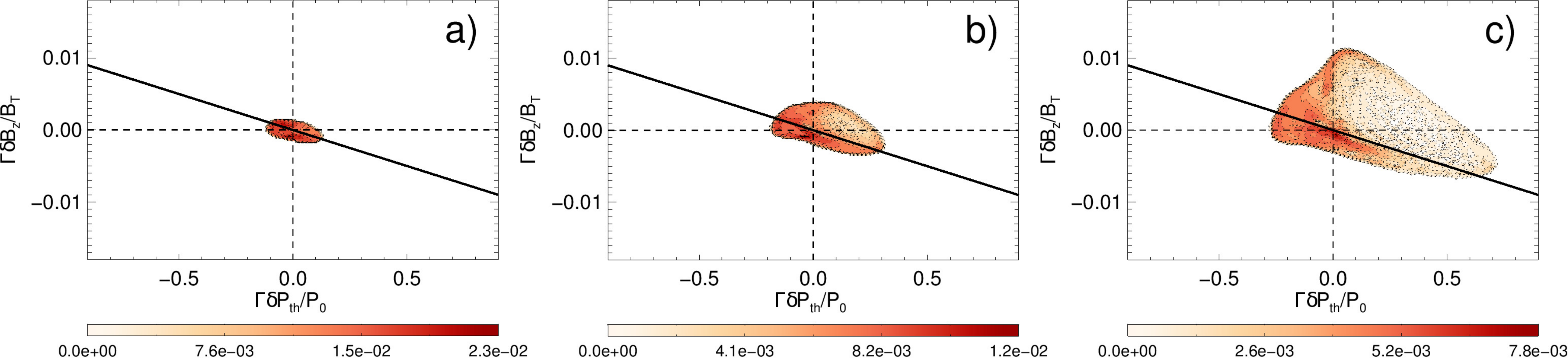}
	\caption{2D (frequency) histograms with the correlation between the magnetic and thermal fluctuations $\Gamma\delta B_z/B_T$ and $\Gamma\delta P_{th}/P_{th, 0}$ for the case PIC $b_g=5$, at the times: a) $t=20\tau_A$, b) $t=30\tau_A$, c) $t=40\tau_A$. See Fig.~\ref{fig:ffbgs-pressurescorrelation}(a2) for the calculation method. \label{fig:evolution_bz_pth}}
\end{figure*}

Although useful, Fig.~\ref{fig:ff-linearscaling} cannot provide us with information about the relation of these maxima with the pressure equilibrium condition Eq.~\eqref{pressure_equilibrium}. For this purpose,  in Fig.~\ref{fig:evolution_bz_pth} we show the 2D histograms relating these fluctuations  in the plane $\delta P_{th}-\delta B_z$  for three characteristic times during the evolution of the case PIC $b_g=5$.
The asymmetry along the pressure equilibrium line located in the right bottom part ($\delta P_{th}>0$ with $\delta B_z<0$) starts well before the reconnection peak time, indicating  a process that is active since the start. Because the points tracked in Fig.~\ref{fig:evolution_bz_pth}(a)  (for $t=20\tau_A$) are mostly located in the separatrices, we will see in Sec.~\ref{sec:bzgeneration} that it is associated with the initial shear flow.
On the other hand,  the ``bump'' $\delta P_{th}\sim 0$ with $\delta B_z\gtrsim0$ (indicating violation of the pressure equilibrium condition and core magnetic field) starts to develop just before of the reconnection peak time ($t\sim30\tau_A$,   Fig.~\ref{fig:evolution_bz_pth}(b)). These points are located inside of the secondary magnetic islands or  the $y$ boundaries. Therefore, and different from the mechanism leading to the asymmetric separatrices, this is an indication of being caused via a process that needs to be build up during the evolution of reconnection. Later,  in  Fig.~\ref{fig:evolution_bz_pth}(c) for $t\sim40\tau_A$,  the points spread even further in the ``bump region'',  indicating the shift of the largest deviations of the pressure equilibrium condition from the separatrices to the ``bump''. In Sec.~\ref{sec:bzgeneration} we will see that the mechanism is a combined effect of both shear flow and magnetic islands generated via magnetic reconnection.
We can summarize this section by saying that PIC and GK simulations predict similar evolution for both magnetic and thermal pressures especially during the linear phase of reconnection. The formation of secondary magnetic islands breaks this similarity, producing deviations especially in the magnetic fluctuations $\delta B_z$,  that reach maxima for times much later than the reconnection peak time.
\section{Core magnetic field and shear flow}\label{sec:bzgeneration}

In this section we describe the physical mechanism that leads to the generation of core magnetic field in the PIC low guide field simulations, complementing  Sec.~\ref{sec:corefield_pressureequilibrium}. In this point, it is important to mention some previous works that have found this feature, such as Ref.~\onlinecite{Zhou2014}. They reported the generation of core magnetic field during the coalescence of secondary magnetic islands, as result of the Hall effect that twists magnetic field lines, plus flux transport with the associated pile up of the out-of-plane magnetic field. In our case, the mechanism is more related to the first one but due to a different reason.
\subsection{Initial shear flow}\label{sec:shear_flow}
The core magnetic field seen in the PIC low guide field runs (see, e.g., Figs.~\ref{fig:bz_contours}(a) and~\ref{fig:bz_contours}(b)) is due to the development of strong in-plane currents growing from  a shear flow present in the PIC initialization.
Indeed, in addition to the component $J_z$ sustaining the CS (associated with $B_y(x)$), the force free initialization with $B_z(x)$ implies, via Amp\`ere's law, the presence of an in-plane component $J_{y}$, chosen to be carried only by the electrons, and given by
{\small
\begin{equation}
	J_{e,y}= \frac{B_{\infty y}}{\mu_0 L} \frac{\displaystyle \left[\frac{\displaystyle\tanh\left(\frac{x-L_x/4}{L}\right)}{\displaystyle\cosh^{2}\left(\frac{x-L_x/4}{L}\right)} + \frac{\displaystyle\tanh\left(\frac{x-3L_x/4}{L}\right)}{\displaystyle\cosh^{2}\left(\frac{x-3L_x/4}{L}\right)}\right]}{\displaystyle \sqrt{b_g^2 + \cosh^{-2}\left(\frac{x-L_x/4}{L}\right) + \cosh^{-2}\left(\frac{x-3L_x/4}{L}\right)}}.
\end{equation}
}

\begin{figure}[h!]
	\centering
	\includegraphics[width=\linewidth]{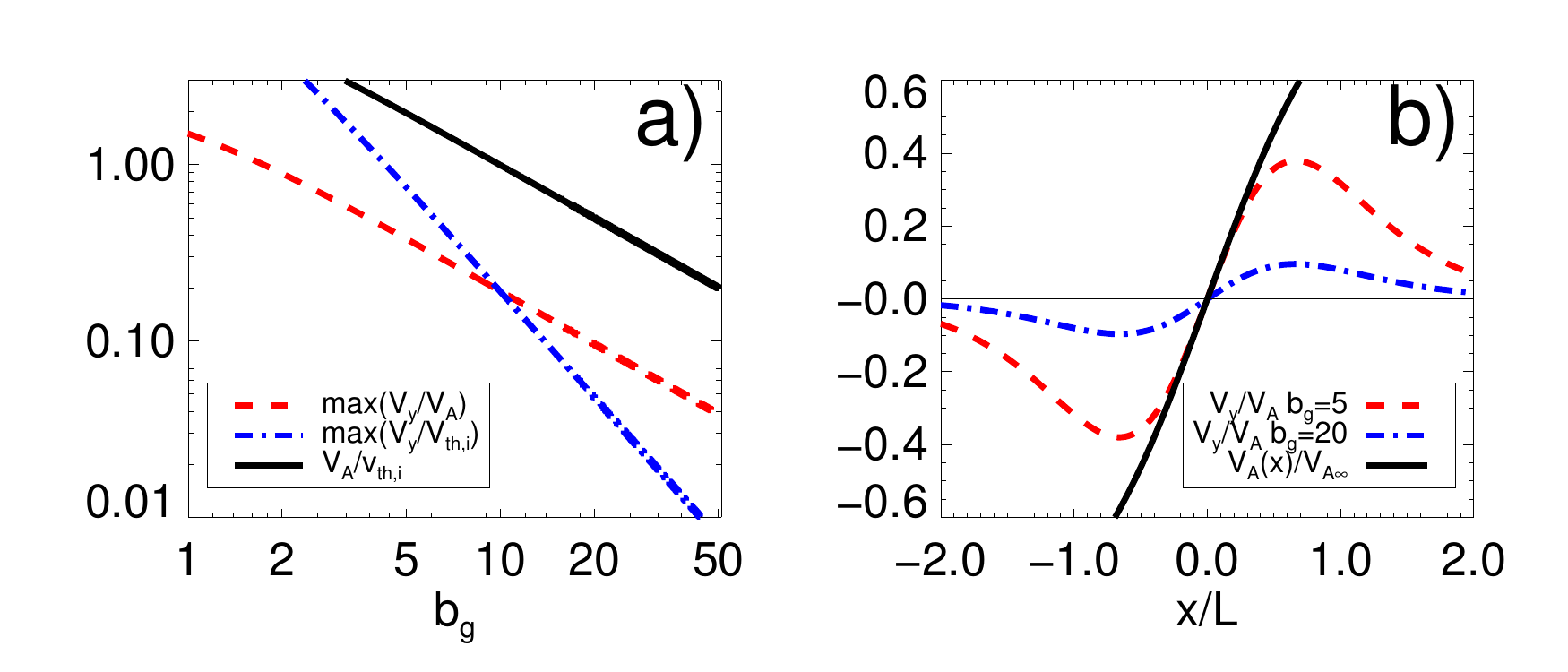}
	\caption{a) Maximum initial value of the  in-plane electron flow speed  $V_{e, y}$ versus $b_g$  normalized to $V_A$ and $v_{th, i}$, as well as the speed ratio $V_{A}/v_{th, i}$ given in Eq.~\eqref{ratio_va_vthi}.\\
		b) Initial profiles of $V_{e, y}(x)/V_{A}$ across a CS centered in $x=0$ for two values of guide field, as well as the local Alfv\'en speed $V_{A}(x)$.\label{fig:multiplot_forcefree_maxjy}}
\end{figure}

This represents a counterstreaming (shear) flow of electrons along the center of each CS (see Fig.~\ref{fig:multiplot_forcefree_maxjy}(b)), with maximum value $V_{e, y}=-J_{e,y}/(en_0)\propto B_{\infty y}$. Thus, due to the normalization used (see  Appendix~\ref{sec:app_normalization}), its magnitude will decrease as $1/b_g$ (see Fig.~\ref{fig:multiplot_forcefree_maxjy}(a)), while the associated kinetic energy of the shear flow has an even stronger dependence ($\propto1/b_g^4$). Therefore, the shear flow strength in the PIC cases converges very quickly to the GK initialization, where it is completely absent.
The zero initial shear flow in this plasma model is because of the perpendicular GK Amp\`ere's law, $(\nabla\times\delta\vec{B})_{\perp}=\nabla_{\perp}\delta B_z = \mu_0 \delta \vec{J}_{\perp}$ (leading to the pressure equilibrium condition Eq.~\eqref{pressure_equilibrium_gk}). Indeed, when $J_{\perp}$ is calculated from the force free particle distribution function, it turns out to be of second order in $\epsilon$. If taken into account, this would imply a $\delta B_z$ of order $\epsilon^2$, which is ruled out by construction from the GK equations. Therefore, any shear flow (in-plane $\delta J_{\perp}$) does not enter into the  GK equations  since they are of order $\epsilon^2$.

The initial shear flow is  responsible for the asymmetric separatrices (see discussion of Figs.~\ref{fig:ffbgs-pressurescorrelation} and \ref{fig:evolution_bz_pth}), especially regarding the  preferential $\delta P_{th}>0$ over $\delta P_{th}<0$. This behavior was already pointed out by \citet{Cassak2011a}, where it was attributed to the dynamic pressure of the shear flow,
piling up electrons preferentially in one pair of the separatrices over the other one if strong enough,  contributing to the increase of density,  temperature and thermal pressure (see Fig.~\ref{fig:pth}).
 An extended discussion about this topic will be given in a follow-up paper.

The effects of the shear flow need to be carefully considered when normalized to $V_A$ (dependent on $b_g$) or $v_{th,i}$ (independent on $b_g$). The ratio between these two speeds is shown in Fig.~\ref{fig:multiplot_forcefree_maxjy}(a), decreasing with $b_g$ according to the normalization used (see Appendix~\ref{sec:app_normalization}). This speed ratio plays an essential role in the GK model,  because of the perpendicular drift approximation. This is the assumption that the perpendicular bulk velocities $\vec{V}_{\perp}$ are caused only by the $\vec{E}\times\vec{B}$, diamagnetic $\nabla P_e$ and other similar drifts. The perpendicular approximation breaks down when the in-plane speeds are close to the ion thermal speeds, such as the typically obtained in magnetic reconnection outflows when $V_A\sim v_{th, i}$. It can also be violated by the presence of waves with high enough frequency (possibly only captured by the fully kinetic plasma description),  leading to a cross-field diffusion of particles across the magnetic field.\cite{Chen2015} Therefore, deviations from the real physical behavior of a Vlasov plasma modeled via PIC simulations are expected in the GK approach when the ratio $V_A/v_{th, i}\sim 1$.  This is precisely the situation when comparing the GK to PIC simulations with $b_g=5$ or $b_g=10$ (the latter is the critical guide field for which $V_A/v_{th, i}\approx 1$).

\subsection{Current/flows in secondary magnetic islands}\label{sec:current_2ndislands}
\begin{figure*}[!ht]
	\centering
	\includegraphics[width=0.8\linewidth]{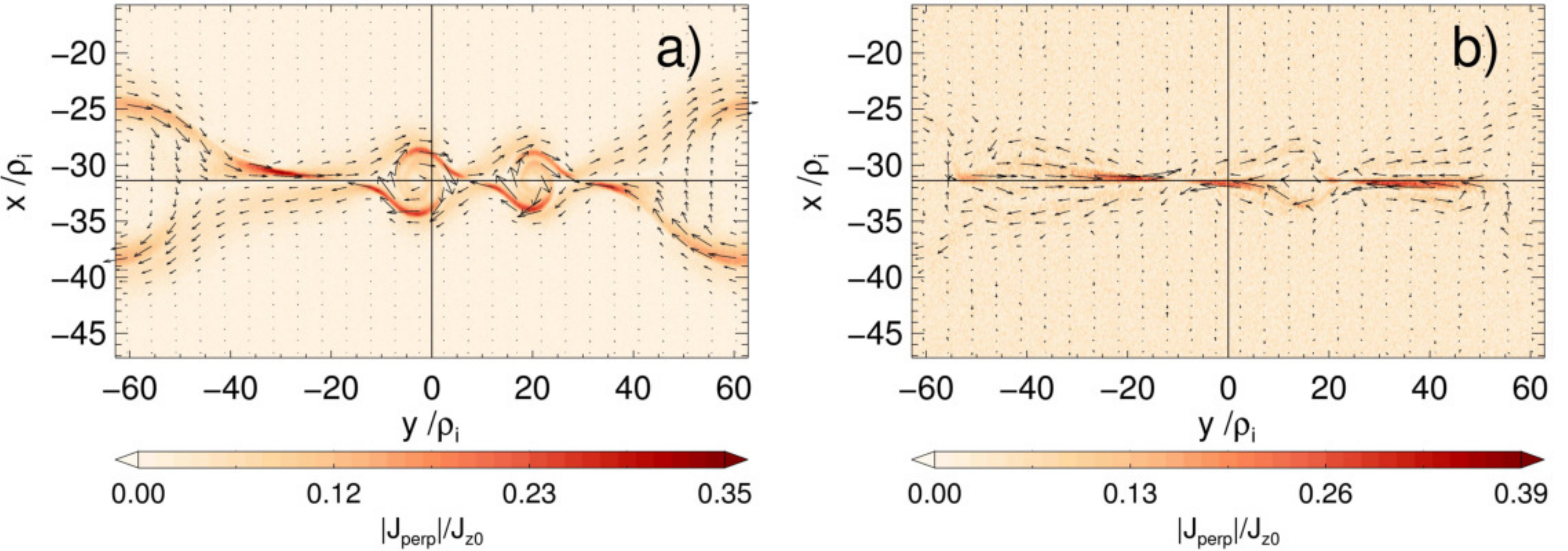}
	\caption{Vector plot of the in-plane current $\vec{J}_{\perp}=J_x\hat{x}+J_y\hat{y}$ for two cases of PIC guide fields a) $b_g=5$ and b) $b_g=20$, at a time $t=50\tau_A$. Color coded is its magnitude $|\vec{J}_{\perp}|/J_z(t=0)$, with $J_z(t=0)$ given in Eq.~\eqref{norma_j}. \label{fig:currents_later}
}
\end{figure*}

As a result of the initial shear flow,  PIC runs with sufficiently low guide field build up a net vortical current inside of the secondary magnetic islands and the periodic $y$ boundaries,  as can be seen in Fig.~\ref{fig:currents_later}. The formation of secondary magnetic islands wraps up magnetic field lines around them, deflecting the electron shear flow in the same direction.  Therefore, a net out-of-plane magnetic field is generated (see Fig.~\ref{fig:bz_contours}). Note that the direction of the curl of $\vec{J}$ coincides with the direction of the out-of-plane magnetic field ($+z$ direction points into the page). As expected, there is a working dynamo process in these places,  i.e.: $\vec{J}\cdot\vec{E}<0$,  involving a transfer of energy from the bulk electron motion to the magnetic field (plots not shown here). High guide field PIC  or GK runs do not show this effect. (Negative values of $\vec{J}\cdot\vec{E}$ are seen only near the outflows, due to the bulk motion of the plasma.)
A comparison of this process to the dissipation  $\vec{J}\cdot\vec{E}>0$ close to the X points will be given in a follow-up paper.

Two other consequences of the shear flow in reconnection found in previous works are worth to mention in this point. First, the tilting of magnetic islands and generation of concentric vortical flows inside of them as result of sub-Alfv\'enic flows (in agreement with our parameter regime) has been seen in both  2D MHD and Hall-MHD simulations (see Ref.~\onlinecite{Shi2005} and references therein).
Second, the increasing symmetry of the core magnetic field for low PIC $b_g$ can also be explained due to the shear flow, as investigated via a two fluid analysis of tearing mode with shear flow of  Ref.~\onlinecite{Hosseinpour2013}. But there is also opposite evidence to this claim: Ref.~\onlinecite{Karimabadi1999} found, by means of hybrid simulations of Harris sheets, that the ``S'' shape of $\delta B_z$ in the secondary magnetic islands is only consequence of guide field reconnection (nothing to do with shear flow). This might be due to the small guide field regime analyzed: $b_g<1$, not being correct to be applied for our case.

\begin{figure*}[!ht]
	\centering
	\includegraphics[width=\linewidth]{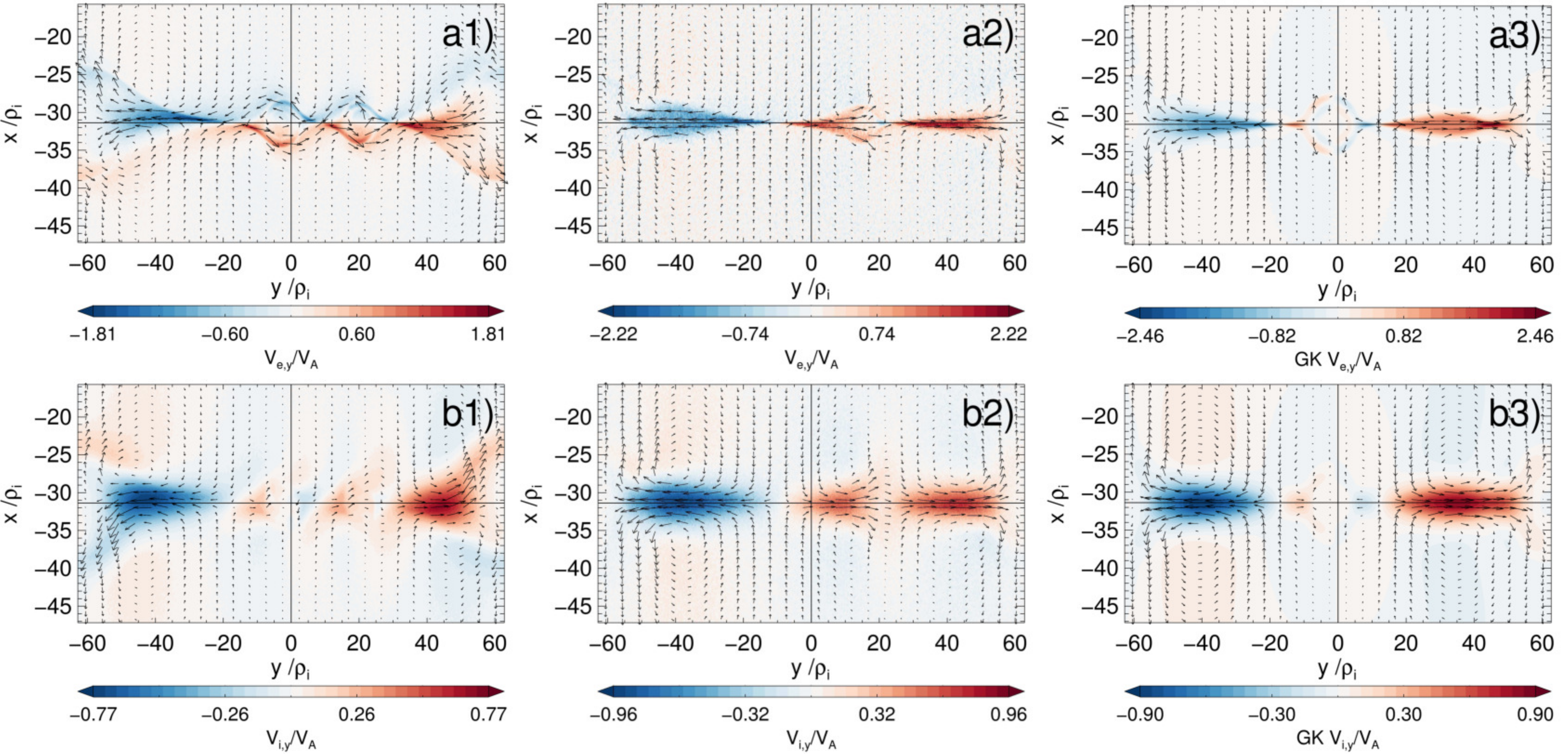}
	\caption{Vector plot of the in-plane bulk velocity $\vec{V}_{e/i, \perp}$ for different PIC guide fields and GK runs, at a time $t/\tau_A=50$. Color coded is the magnitude $|\vec{V}_{e/i, y}/V_A|.$ Guide field increases to the right:  ax) PIC $b_g=5$, bx) PIC $b_g=20$, cx)  GK, where $x=1$ (top row) is for the electron $\vec{V}_{e, \perp}$ and $x=1$ (bottom row) is for the ion $\vec{V}_{i, \perp}$.\label{fig:elec_ion_flows}}
\end{figure*}

The current inside of the magnetic islands,  significant only in the low guide field regime,  is produced due to the Hall effect: a decoupling of motion between electrons and ions,  as shown in Fig.~\ref{fig:elec_ion_flows}.
For the lowest PIC guide field run $b_g=5$,  the ions follow the outflow due to reconnection from the X point (Fig.~\ref{fig:elec_ion_flows}(b1)),  but the electrons keep their initial shear flow and are only weakly deflected (Fig.~\ref{fig:elec_ion_flows}(a1)) by that reconnection outflow,  following a vortical flow pattern inside of the magnetic islands. This characteristic (antisymmetric) flow pattern is barely visible for a higher guide field of $b_g=20$ (see Figs.~\ref{fig:elec_ion_flows}(a2)-(b2)) and totally absent for the GK run (see Figs.~\ref{fig:elec_ion_flows}(a3)-(b3)), where the internal flow has a symmetric structure following the features of the reconnection outflow.

It is important to mention that because the generation of magnetic field is due to a Hall effect,  their effects will be stronger when the CS is on the order of/thinner than $\rho_s=\sqrt{k_B(T_e+T_i)/m_i}/\Omega_{ci}=\rho_i=0.1d_i$, the sound Larmor radius (for $T_i=T_e$),
which applies very well to our case ($L=2\rho_i$). Therefore, in thicker CS these effects should be significantly reduced and an agreement between PIC and GK simulations of magnetic reconnection will be more easily reached. Note that the halfwidth is also on the order of electron skin depth, since $\rho_i=d_e$, and so the electron inertial effects are important:\cite{Nakamura2008} electrons are also unmagnetized (not fulfilling the frozen-in condition) for distances $l\lesssim L/2$.

In Fig.~\ref{fig:elec_ion_flows}, we can estimate that the speed of the electron outflows from the main X point is a few times the asymptotic Alfv\'en speed $V_{e, y}\sim 2.2 V_A$. It varies weakly with the guide field  in the PIC runs.  On the other hand, the ion outflow speeds only reach sub-Alfv\'enic values  $V_{i, y}\sim 0.8 V_A$, and are much more constant among different PIC guide fields and the GK runs. These values agree,  to a first order,  with a two fluid theory of magnetic reconnection by \citet{Shay2001b}: the outflow speeds from the X point should be of the order of the in-plane Alfv\'en speed $V_{A}$ for ions and in-plane electron Alfv\'en speed $V_{Ae}=\sqrt{m_i/m_e}V_{A}$  for electrons. But the initial shear flow is of course strongly dependent on the guide field (see Fig.~\ref{fig:multiplot_forcefree_maxjy}). The combination of both effects determines the critical PIC guide field for which the shear flow can generate the current that builds up the core magnetic field ($b_g\lesssim 20$).

\subsection{Time evolution of the current and magnetic field generation}
\begin{SCfigure*}[\sidecaptionrelwidth][!ht]
	\centering
	\includegraphics[width=1.4\linewidth]{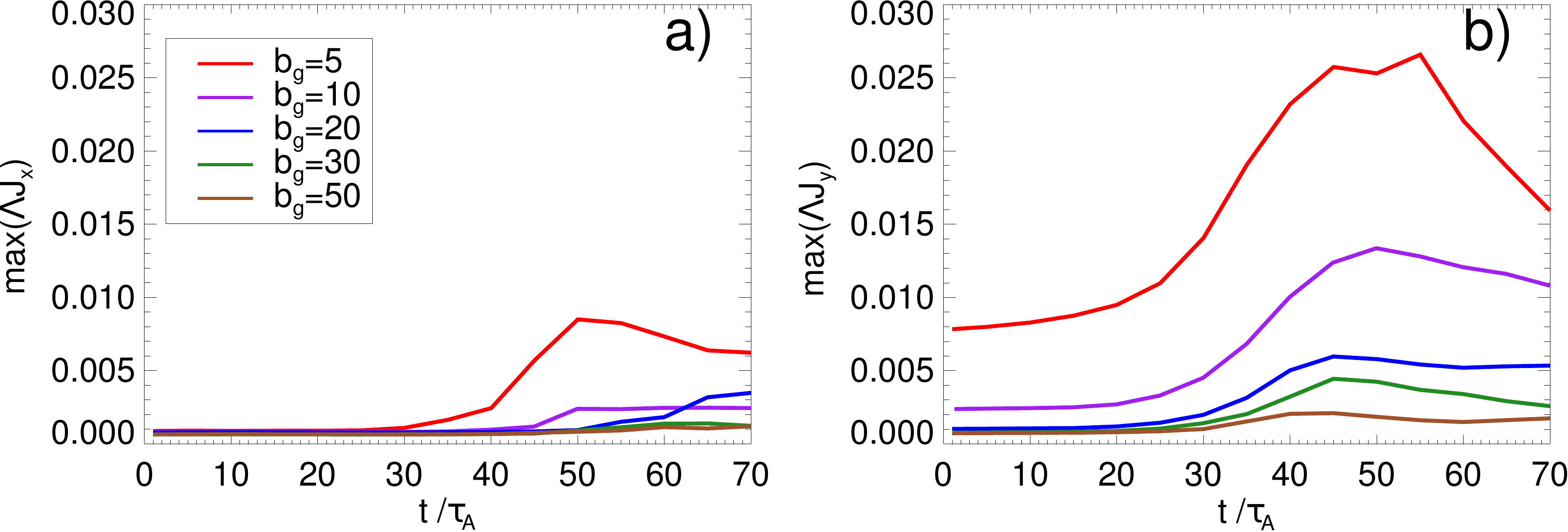}
	\caption{Time history of the maximum value of the in-plane current density for different PIC guide fields cases. a) $\Lambda J_{x}$. b) $\Lambda J_{y}$. Note the absolute units (no normalization depending on $b_g$ like in Fig.~\ref{fig:currents_later}).  These values have to be compared with the ones for $\delta B_z/B_g$ shown in Fig.~\ref{fig:ff-linearscaling}.\label{fig:evolution_j}}
\end{SCfigure*}

Since in Sec.~\ref{sec:current_2ndislands} we showed that the electron/ion outflows do not depend strongly on $b_g$ when normalized to $V_A$,
the in-plane current $\vec{J}_{\perp}\propto\vec{V}_{i, \perp} - \vec{V}_{e, \perp}$ will  display similar values among different PIC $b_g$ when the same normalization is used, as can be seen in Fig.~\ref{fig:currents_later}. But according to the Amp\`ere's law, the generation of $\delta B_z$ depends on the unnormalized $J_{\perp}$, changing with $b_g$ according to Eq.~\eqref{ratio_va_vthi}. It can be estimated  by approximating the curl $\nabla\times\vec{B}$ by the gradient scale length $1/\Delta L$:
\begin{align}\label{bz_estimation}
	\frac{\delta B_z}{B_g}\approx \frac{\Delta L}{\rho_i}\left(\frac{\mu_0 \rho_i}{B_g}\right)J_{\perp}.
\end{align}
Since $B_g$ practically does not change for different PIC guide fields $b_g$ (result of choosing an invariant $B_T$, recall Sec.~\ref{sec:setup}), by defining the constant $\Lambda=\mu_0 \rho_i/B_g$  we infer that (in order of magnitude), $\delta B_z/B_g\sim \Lambda J_{\perp}$ when $\Delta L\sim\rho_i$,  not dependent on the guide field. Note that this estimate relies on the fact that $\rho_i$ is approximately constant during the evolution, which in turn depends on the fact that $T_i$ does not have to change too much. We checked that the ion temperature does not increase more than $30\%$ of its initial value throughout the evolution of the system, and therefore we can safely take $\rho_i$ and so $\Lambda$ constant for our purposes. This is equivalent to a proportionality between $\delta B_z$ and $J_{\perp}$, a relation always satisfied in a force free equilibrium.  This is a valid first order approximation since the magnetic islands grow at ion time scales ($\tau_A$), and therefore, at electron time-scales that relation can be satisfied quasi-adiabatically. On the other hand, in our runs  $\Delta L$ can be estimated as the size across the $x$ direction of either the secondary magnetic islands close to the $X$ point ($\sim 10\rho_i$), or the magnetic island at the $y$ boundaries ($\sim 18\rho_i$). Thus, the time history of the maximum value of both components of $\Lambda J_{\perp}$ is shown in Fig.~\ref{fig:evolution_j}.

The presence of the initial $J_y$ due to the shear flow can be seen in Fig.~\ref{fig:evolution_j}(b), increasingly important for lower PIC guide fields. But the in-plane component $J_y$, and to a lesser extent $J_x$, start to grow just after the formation of secondary magnetic islands.
There is a good agreement (in order of magnitude) of the maximum values that $\Lambda J_{y}$  reach with the corresponding maximum values of $\delta B_z$ (see Fig.~\ref{fig:ff-linearscaling}(b)) for different guide fields, justifying empirically our estimation Eq.~\eqref{bz_estimation}. Note that the factor $\Gamma$ should be removed in the latter for a proper comparison (e.g.,  for $b_g=5$,  the maximum is $\delta B_z/B_g\sim 0.034$). It is also important to remark that the maximum values of $J_y$ are reached in the boundaries of the locations where $\delta B_z$ peaks: around the secondary magnetic islands and the $y$ boundaries (neglecting the region in the separatrices with almost zero curl). The most important conclusion that can be inferred from Fig.~\ref{fig:evolution_j} is that the maximum values of $J_y$,  and so $\delta B_z$, become negligible in the PIC higher guide field limit when measured in absolute units. Therefore,  magnetic field generation is only effective in the PIC low guide field regime.

In principle, another possibility for the core-magnetic field generation that it is interesting to analyze is due to a pinch effect and the associated magnetic flux compression inside of the magnetic islands (via conservation of magnetic flux/frozen-in condition). We checked that the divergence $\nabla \cdot \vec{v}_{\perp}$ for both electrons and ions has a complex spatial structure and mostly odd symmetry inside and around the islands, not correlating well with the properties of $\delta B_z$ (plots not shown here). Moreover, we checked that the absolute value of $\nabla \cdot \vec{v}_{\perp}$ (normalized to the natural time scale $\tau_A^{-1}$), is reduced for higher guide fields, implying a more incompressible plasma. This is consequence of the fact that the lowest order perpendicular drifts are incompressible, a condition  satisfied better in this regime. However, $\nabla \cdot \vec{v}_{\perp}$ decays much faster with the guide field as the core-field: already for $b_g=20$, this quantity is comparable with the noise level but, on the other hand, there is still significant $\delta B_z$ in the islands for this guide field case.  Therefore, a pinch effect $\nabla \cdot \vec{v}_{\perp}<0$ is not likely to be responsible for the generation of $\delta B_z$ in the magnetic islands.

All these are indications that the generation of the core magnetic field $\delta B_z$ is due to a combination of the effects of the shear flow and the formation of magnetic islands, becoming increasingly important for lower PIC guide fields, and that is why it should not be taken into account when comparing PIC with GK simulations of magnetic reconnection.
\subsection{Effects of counterstreaming outflows in core magnetic field generation}\label{sec:numerical_bz_generation}
We already mentioned that in the PIC low guide regime,  besides of the secondary magnetic islands,  a core magnetic field is also generated in the $y$ boundaries. This is due to a combination of two mechanisms. The main one is because of the compression  by the colliding outflows generated from reconnection in the main X point and the periodic boundary conditions (equivalent to a configuration with multiple X points). This effect is stronger for PIC low guide field regime since the electron outflows are faster (in absolute units, since they have same values in units of $V_A$). Note that this process could be avoided by choosing longer boxes.\cite{Karimabadi1999}  The second mechanism contributing to the core magnetic field generation is the same as for the secondary magnetic islands, since the O point of the primary magnetic island is located precisely at the  $y$ boundaries.

\subsection{Influence of shear flow in reconnection}
As we can see in Fig.~\ref{fig:recrates}, the reconnection rates are smaller for the lower guide field cases $b_g=5$ and $10$ compared to the PIC high guide field regime or the GK runs. This can be understood in terms of three different reasons. First,  the outflows from the X points should be less efficient (slower) because of the additional magnetic pressure due to $\delta B_z$ in the secondary magnetic islands (and $y$ boundaries), after being normalized to the Alfv\'en speed. This process tends to inhibit the CS thinning. Note that this additional out-of-plane guide field has higher relative importance with respect to the asymptotic magnetic field precisely in cases with low PIC $b_g$ (see,  e.g., Fig.~\ref{fig:ff-linearscaling}). Second,  as indicated in Ref.~\onlinecite{Cassak2011a} with the Hall-MHD model without guide field, the magnetic tension in the reconnected magnetic field lines should be released by the shear flow, reducing the outflow speed produced by them and thus decreasing reconnection rates by a factor of $\left( 1 -V_{e,y}^2/V_A^2\right)$. Similarly, numerical solutions of a kinetic dispersion relation and 2D PIC simulations for thin CS\cite{Roytershteyn2008} demonstrated a reduction in the growth rate of the collisionless tearing mode  for increasing shear flows, the instability associated with the onset of magnetic reconnection. And third, part of the available magnetic energy for reconnection that should be converted into particle energy is given back when the Hall currents are formed in the secondary magnetic islands.
 Overall, these reasons suggest that reconnection rates might be reduced in the PIC low guide field regime,  if the total plasma $\beta$ is kept constant.

\section{Finite plasma beta effects}\label{sec:finite_plasmabeta}
\begin{SCfigure*}[\sidecaptionrelwidth][!ht]
	\centering
	\includegraphics[width=1.4\linewidth]{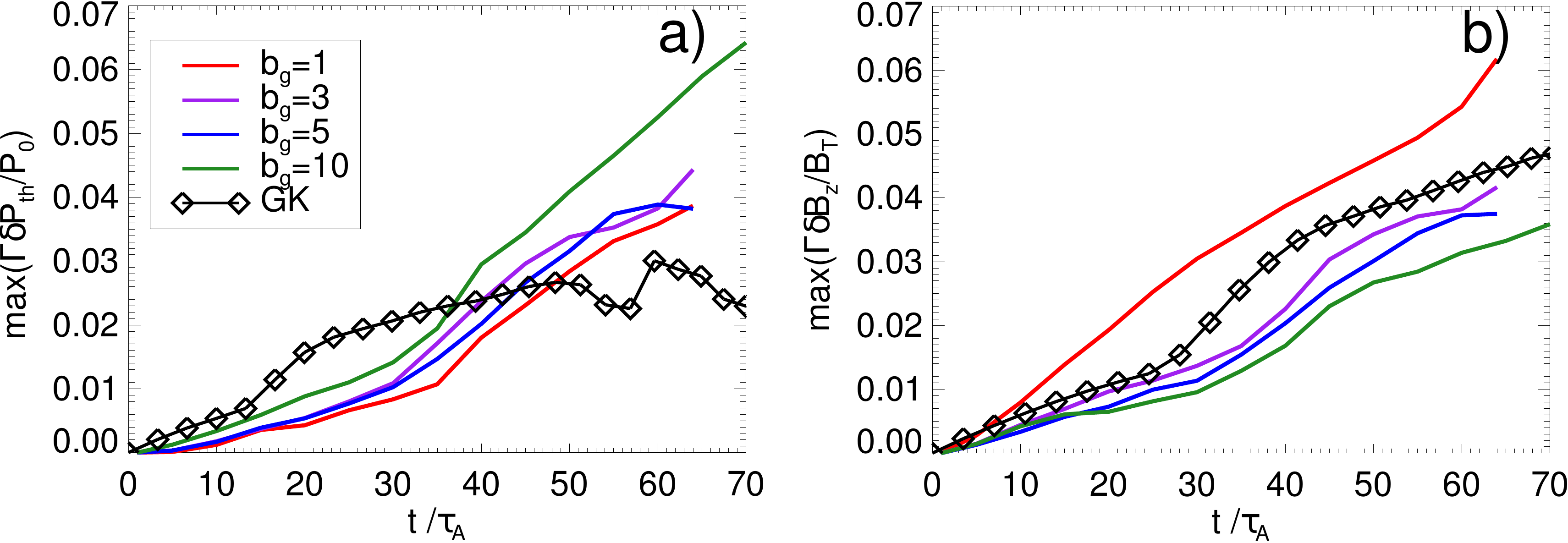}
	\caption{Same time histories as Fig.~\ref{fig:ff-linearscaling} but for the high beta case. (a): scaled thermal pressure $\Gamma\delta P_{th}/P_{th, 0}$ and (b) magnetic fluctuations $\Gamma\delta B_z/B_T$. $\Gamma$ is for $b_{g,ref}=5$ in  Eq.~\eqref{scaling_factor}. \label{fig:finite_beta_effects}}
\end{SCfigure*}

Let us finally discuss the high beta case with $\beta_i=1.0$. Although the basic phenomenology of magnetic reconnection and secondary magnetic islands is similar, in general, the agreement is better between different guide field PIC runs in the range $b_g=1\to10$ and the corresponding GK results. This  can be seen in the reconnection rates of Fig.~\ref{fig:recrates}(b),  as well as in the time evolution of the scaled magnetic fluctuations $\Gamma\delta B_{z}$ shown in Fig.~\ref{fig:finite_beta_effects}(b). The time evolution of the thermal pressure fluctuations $\Gamma\delta P_{th}$ in Fig.~\ref{fig:finite_beta_effects}(a) displays a different behavior between the results given by both codes.
As we already pointed out,  the physical origin of these differences will be analyzed in a follow-up paper.

 There are at least two related reasons for the better agreement. The first one is because the fluctuations $\delta P_{th}/P_{th,0}$  are predicted to be smaller by one order of magnitude compared to the low beta case, since they are proportional to $\propto 1/\sqrt{\beta_i}$ (see Eqs.~\eqref{delta_thermalpressure} and \eqref{delta_magneticpressure}). In addition, the GK ordering parameter $\epsilon$ will also be reduced (see discussion of Eq.~\eqref{epsilon_gk}), from $\epsilon=4$ for $b_g=5$ in the low beta case to  $\epsilon=0.2$ for $b_g=1$ in the high beta case. This improves the validity of the predictions of the GK compared to the PIC model in this parameter regime, since the largest deviations from the pressure equilibrium condition Eq.~\eqref{pressure_equilibrium} will be reduced by the same amount.
Also note that the maximum net force due to the total pressure imbalance will be much smaller than in the low beta case.

\begin{figure*}[!ht]
\centering
\includegraphics[width=\linewidth]{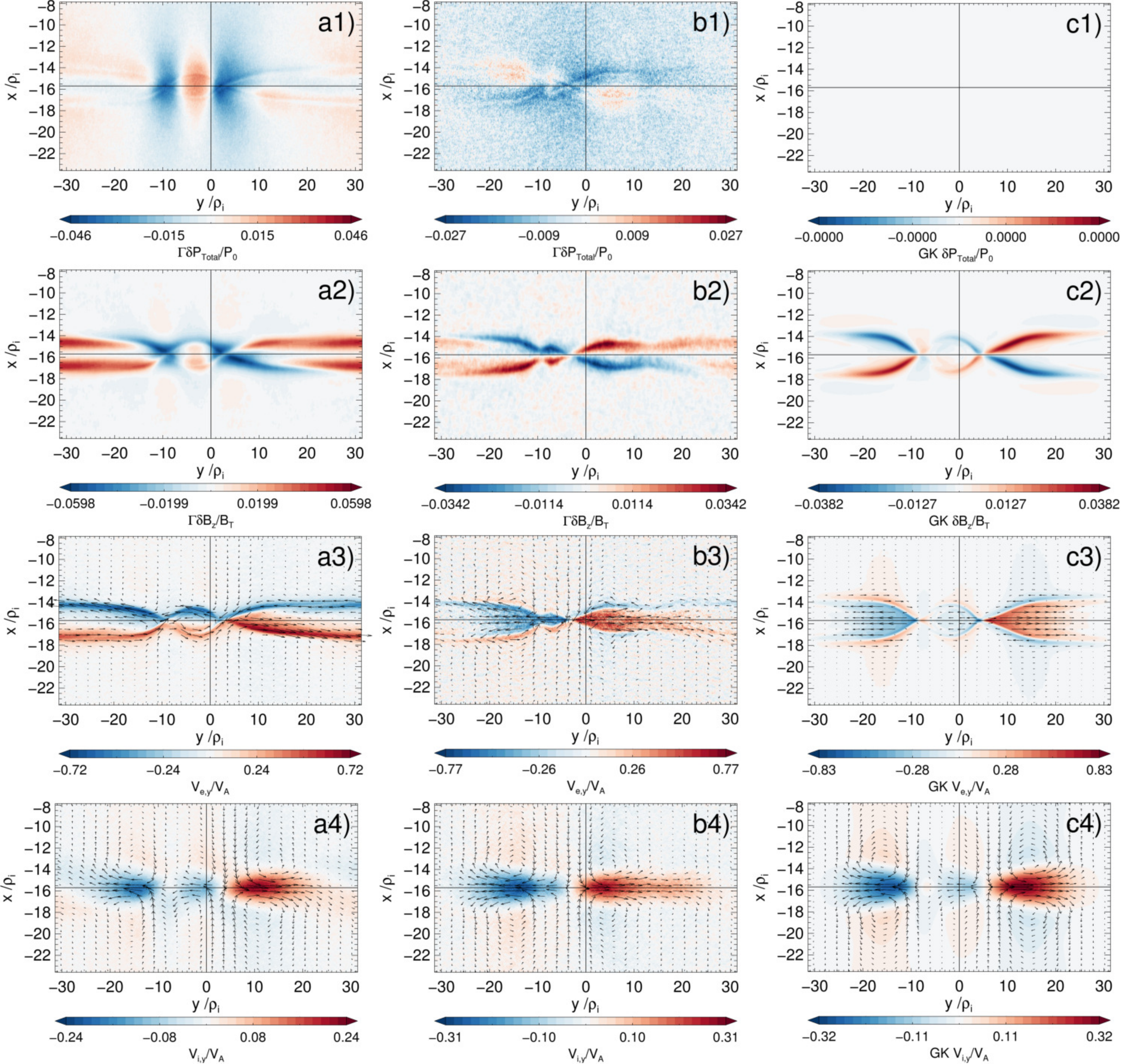}
 \caption{
Contour plots showing some quantities in the high beta case $\beta_i=1.0$  for two PIC guide fields and GK runs, at a time $t=50\tau_A$. Guide field increases to the right:  ax) PIC $b_g=1$, bx) PIC $b_g=5$, cx) GK. Row $x=1$: scaled total pressure $\Gamma\delta P_{\rm total}/P_{th, 0}$. Row $x=2$: scaled magnetic fluctuations $\Gamma\delta B_z/B_T$. Row $x=3$: Vector plot of the in-plane electron bulk velocity with color coded the component $V_{e, y}/V_A$. Row $x=4$:  Vector plot of the in-plane ion bulk velocity with color coded the component $V_{i, y}/V_A$.
\label{fig:highbeta_contours}}
 \end{figure*}

 Second, the range of variation of the speed ratio given by Eq.~\eqref{ratio_va_vthi} decreases by a small amount in comparison to the low beta case: $V_A/v_{th,i}=0.1\to 0.7$ going from  $b_g=10\to 1$. This implies that the maximum values of the initial shear flow (for $b_g=1$) are much smaller: $\max(V_{y0})\sim 0.3 V_A \sim 0.2v_{th, i}$, and so are their effects on the system.
 Moreover,  reconnection rates are reduced by a factor of two in this regime (see Fig.~\ref{fig:recrates}(b)). Then,  the maximum electron/ion outflows speeds in units of $V_A$,  proportional to the reconnection rates,  will also be further reduced in comparison to the low beta case,  becoming negligible when measured in units of $v_{th, i}$. More precisely,  we measured maximum electron outflows speeds on the order of $0.75V_A$ (see Fig.\ref{fig:highbeta_contours}(3rd. row)) and ion outflows on the order of  $0.3V_A$ (see Fig.\ref{fig:highbeta_contours}(4th. row)),  roughly smaller by a factor of 3 compared with the corresponding values in the low beta case.

 This reduction in the maximum in-plane flow speeds has two physical consequences. For the GK model, the maximum deviations from the drift approximation
 will be smaller than in the low beta case (see discussion in Sec.~\ref{sec:shear_flow}).
For the fully kinetic approach, the magnetic field generation due to the colliding outflows at the $y$ boundaries is reduced.
On the other hand, we also observed smaller magnetic islands in this case, implying a smaller core-magnetic field. This is because the generation of $\delta B_z$,  according to the estimate Eq.~\eqref{bz_estimation},  is proportional to the length scale $\Delta L$ of these islands,  which is roughly five times smaller compared to the low beta case. This is valid even though the Hall term and the corresponding decoupling of electrons and ions is facilitated in high plasma beta environments, implying a greater in-plane $J_{\perp}$ (twice as high compared to the low beta case).
The net result of all these effects can be seen in the time evolution of $\Gamma \delta B_z$ in Fig.~\ref{fig:finite_beta_effects}(b),  where the deviations of PIC results compared to the GK ones are significant only for $b_g=1$ and much smaller in absolute terms compared with the low beta case. Convergence with the GK results is already reached with values $b_g\gtrsim 3$ (see, e.g., the second column of Fig.~\ref{fig:highbeta_contours} for $b_g=5$). This reduction of core magnetic field strength in high $\beta$ plasmas is in agreement with previous hybrid simulations\cite{Karimabadi1999} (although for very low guide fields $b_g<1$).

Finally, it is important to mention that there is, in general, a higher level of numerical noise,  and associated numerical heating,  in this high beta case compared to the low beta one for the PIC runs. It becomes increasingly important for higher guide fields,  reducing even faster the signal-to-noise ratio. This is,  in part,  responsible for the monotonically increasing thermal pressure fluctuations in Fig.~\ref{fig:finite_beta_effects}(a) for later times,  especially in the case $b_g=10$. This observation was already pointed out in Ref.~\onlinecite{TenBarge2014}, being a well known consequence of the enhanced numerical collisions in weakly magnetized environments simulated by PIC codes,  or equivalently, in high beta plasmas.\cite{Matsuda1975}

\section{Conclusions}\label{sec:conclusions}
We have carried out a comparison of magnetic reconnection  in the limit of strong guide field between two plasma models by using fully kinetic PIC with gyrokinetic simulations of force free current sheets. Our study extends the previous work of Ref.~\onlinecite{TenBarge2014}. We established the limits of applicability of the gyrokinetic approach compared to the fully kinetic model in the realistic regime of finite PIC guide fields, and the physical reasons behind these differences. Note that the following conclusions are based on sets of runs with total ion plasma $\beta_i=0.01$ constant for different PIC guide fields.

First,  by using an independent set of PIC and gyrokinetic codes, we found the limitations in the linear scaling reported in Ref.~\onlinecite{TenBarge2014}  in both thermal and magnetic fluctuations. PIC simulations in the low guide field regime show deviations in the inverse scaling with the guide field, not converging properly to the gyrokinetic results. These deviations start to be especially significant when secondary magnetic islands start to form, after the reconnection peak time.
In this paper, we focus on the additional magnetic fluctuations  revealed by the PIC low guide field runs, which are mostly due to macroscopic bulk plasma motions. The analysis of the differences in the thermal fluctuations---that have  to do with dissipative processes, heating mechanisms and non thermal effects---will be addressed in a follow-up paper.

In particular,  we found that the PIC low guide field regime shows an excess of magnetic pressure inside of the secondary magnetic islands (core magnetic field), reaching maximum values for times much later than the reconnection peak time. Correspondingly, the agreement between the two plasma descriptions is better before the formation of these structures, during the linear phase of magnetic reconnection. This excess of magnetic pressure is  not compensated by a corresponding decrease in the thermal pressure.
The reason is that gyrokinetic codes keep the pressure equilibrium condition to machine precision because perpendicular force balance is enforced in the GK equations,  while PIC codes allow large deviations from it since they solve the full collisionless Vlasov-Maxwell system of equations. Therefore, although the gyrokinetic results can be comparable to the PIC ones for relatively low guide fields ($b_g=5$ and $10$) in the separatrices close to the X points, convergence in the secondary magnetic islands requires much higher guide fields ($b_g \gtrsim 30$).

We found that the physical mechanism that generates the core magnetic field (associated with a dynamo effect $\vec{J}\cdot\vec{E}<0$) is due to an initial shear flow present in the force free PIC initialization.  This shear flow, that also produces asymmetric separatrices and reduces reconnection rates, is negligible in the limit of strong guide field and absent in the corresponding gyrokinetic initialization. Through the Hall effect that decouples electron from ion motion, vortical electron flow patterns are generated in the secondary magnetic islands as magnetic reconnection wraps up magnetic field lines around these structures,  carrying the electron shear flow with them. This magnetic field weakens for higher PIC guide field runs,  converging to the gyrokinetic result,  since the shear flow is not strong enough to drive the currents capable of generating it (when measured in absolute units). There is also an out-of-plane magnetic field generation at the periodic $y$ boundaries, where it is located at the O point of the primary magnetic island. This is not only due to the same previous mechanism, but also due to the compression by colliding electron outflows, faster in the PIC runs with low guide field (in absolute units).

We also showed that the relative ratio of the electron outflow speed to the ion thermal speed, proportional to $V_A/v_{th,i}$, is higher for the PIC low guide field regime, reaching values close to 1 for $b_g\lesssim 10$. This breaks the perpendicular drift approximation,  a critical assumption in which the gyrokinetic codes (and the gyrokinetic theory in general) are based, and thus an additional source of differences is expected compared to the PIC simulation results.

Finally,  we also analyzed the effects of a high plasma beta $\beta_i=1.0$ compared to our standard case $\beta_i=0.01$. Although the basic phenomenology is similar, we could notice a better agreement between the corresponding PIC and gyrokinetic results, reaching a relatively good convergence for guide fields as low as $b_g \sim 3$. From this, we can conclude that an accurate comparison between PIC and gyrokinetic force free simulations of magnetic reconnection requires high plasma $\beta\sim 1$ (to reduce the fluctuation level  proportional to $1/\sqrt{\beta_i}$),  although PIC codes are affected by enhanced numerical heating in this regime. Moreover, it is necessary to have parameters with low ratios $V_A/v_{th,i}\ll1$ (to avoid the effects of the initial shear flow), a  small ordering parameter $\epsilon\ll1$ in the gyrokinetic initialization, and reconnection rates should not be too high $(d\Psi/dt)/\dot{\Psi}_N\lesssim 0.1$. In the PIC model, this is to avoid super-Alfv\'enic outflows generating stronger magnetic fields at the periodic boundaries, while in gyrokinetic model these flows might also break the perpendicular drift approximation.
The balance of these related numerical and physical parameters allow to determine a convenient parameter regime for an accurate comparison of magnetic reconnection between these plasma models.

In spite of all those differences, the gyrokinetic simulations show a development of magnetic reconnection remarkably similar to the PIC high guide field runs. This is of central importance due to the computational savings of the first ones. Indeed,  for the low beta case,  we measured speed-ups by a factor of $10^3$ between the gyrokinetic runs compared to a corresponding PIC guide field $b_g=50$ simulation. However,  the computational cost of a PIC simulation decreases linearly with the guide field, in such a way that in the low guide field regime ($b_g\sim 5$),  the speed-up and advantages of using gyrokinetic instead of PIC simulations are not so notorious.
\begin{acknowledgments}
	We acknowledge the developers of the code ACRONYM at W\"urzburg University, the support by the Max-Planck-Princeton Center for Plasma Physics and of the German Science Foundation DFG-CRC 963 for P. K and J. B.  P. M. acknowledges the financial support of the Max-Planck Society via the IMPRS for Solar System Research. The research leading to these results has received funding from the European Research Council under the European Union's Seventh Framework Programme (FP7/2007-2013)/ERC Grant Agreement No. 277870.

		All authors thank the referee for valuable and constructive comments which helped us to verify our results and to make our points more clear.
\end{acknowledgments}
\appendix
\section{Normalizations}\label{sec:app_normalization}

Here, we explained all the details regarding normalization and the right choice of a correspondence between the GK and PIC results.

 The lengths are normalized to $\rho_i$ and the times to the Alfv\'en time $\tau_A=L/V_A$, where $V_A$ is the Alfv\'en speed defined with respect to the reconnected magnetic field
\begin{equation}\label{alfven_speed}
	\frac{V_A}{c} = \frac{1}{c}\frac{B_{\infty y}}{\sqrt{\mu_0n_0m_i}} = \frac{1}{\sqrt{1+b_g^2}(\omega_{pe}/\Omega_{ce})\sqrt{m_i/m_e}},
\end{equation}
where $\omega_{pe}=\sqrt{n_e e^2/(\varepsilon_0 m_e)}$ is the electron plasma frequency.
We use the previous definitions for the normalization of the reconnection rate: $\dot{\psi}_N=B_{\infty y}V_A$ and current density: $J_N=en_0V_A\sqrt{\beta_i}=en_0v_{th, i}/\sqrt{1+b_g^2}$. Note that due to the Amp\`ere's law, the normalized initial out-of-plane current density that supports the asymptotic magnetic field $B_{y}$ is given by:
\begin{align}\label{norma_j}
J_{zN}:=\frac{J_z(t=0)}{J_N}=\frac{en_0U_e}{J_N}=\frac{U_e}{v_{th,i}}\sqrt{1+b_g^2}=\frac{1}{(L/d_i)\sqrt{\beta_i}},
\end{align}
which is independent on the guide field strength.

As we mentioned in Sec.~\ref{sec:symmetry}, many of the fluctuating quantities scale with the guide field due to the fact that $\beta_i$, and so the transverse distance $l_x/d_i$,  are constant for different $b_g$ (with the last claim proved in Appendix~\ref{sec:app_transverse_distance}). Therefore, the PIC fluctuating quantities will have the same value under the previous assumption if they are multiplied by the  factor proportional to $b_g$
\begin{equation}\label{scaling_factor}
	\Gamma=\frac{\sqrt{1+b_g^2}}{b_{g, ref}},
\end{equation}
where $b_{g, ref}$ is a reference guide field,  chosen to be $b_g=10/ 5$ in the low/high beta case,  respectively.  The value $b_g=10$ in the low beta case was chosen because the respective runs are not so dominated by numerical noise as for larger guide fields ($b_g\gtrsim 30$) and so it is easier to compare with the noiseless GK runs. In addition, $b_{g, ref}=10$ is not in the lowest end of guide fields like $b_g=5$ where other effects not captured by GK (explained in the results, Secs.~\ref{sec:linearscaling}, \ref{sec:corefield_pressureequilibrium} and \ref{sec:bzgeneration}), dominate the physics of the system.   In this way, using Eqs.~\eqref{scaling_factor} and \eqref{delta_thermalpressure},  the quantity that should be equal among different guide field PIC runs is
\begin{align}
	\Gamma\delta P_{th}=\frac{\sqrt{1+b_g^2}}{b_{g, ref}}\delta P_{th}=\frac{\sqrt{1+b_g^2}}{10}\delta P_{th},
\end{align}
and analogously for $\delta B_z$. The factor $\sqrt{1+b_g^2}$ instead of $b_g$ is in order to keep the total magnetic field $B_T$ constant (and not only $B_g$).

On the other hand, the definition of $\Gamma$ in Eq.~\eqref{scaling_factor} also fixes the  ordering parameter $\epsilon$ of the GK runs,  defined by:
\begin{equation}\label{epsilon_gk}
	\epsilon=\frac{1}{b_{\infty y,  norm}b_{g, ref}},
\end{equation}
where $b_{\infty y,  norm}= B_{\infty y}/B_{\infty y,  ref}$ is the normalized asymptotic magnetic field with respect to a reference value $B_{\infty y,  ref}$ expressed in code units. The initialization in the GK runs gives $b_{\infty y,  norm}=0.05/2.5$ for the low/high beta cases,  respectively. Therefore, using the aforementioned PIC $b_{g, ref}$, we have $\epsilon=2/0.04$  for the low/high beta case, respectively. Note that even though in the low beta plasma regime with $\epsilon>1$  the GK ordering formally does not hold, and in agreement with Ref.~\onlinecite{TenBarge2014}, this plasma model can still make accurate predictions, but only under some circumstances (clarified in the results, Secs.~\ref{sec:linearscaling}, \ref{sec:corefield_pressureequilibrium} and \ref{sec:bzgeneration}). Moreover, the unusually high value of $\epsilon$ is required to have a perturbed current strong enough to generate the relative large perturbed asymptotic magnetic field for the case $b_g=10$ (being reduced for higher guide fields).

The choice of a constant $\beta_i$ for different guide fields has another important consequence for the ratio of Alfv\'en to ion thermal speed. Indeed, due to that assumption,  the PIC runs will invariably have to change this parameter for different guide fields
\begin{align}\label{ratio_va_vthi}
	\frac{V_A}{v_{th, i}}=\frac{1}{\sqrt{1 + b_g^2}}\frac{1}{\sqrt{\beta_i}}.
\end{align}
Note that $v_{th, i}$ is constant for different guide fields,  while $V_A$ scales inversely with it. This means that the ratio $V_A/v_{th, i}$ increases for lower PIC guide fields: from $V_A/v_{th, i}=0.2\to1.96 $ going from  $b_g=50\to 5$ in the low beta case.

It is also important to mention another side effect of the normalization used (already noticed in Ref.~\onlinecite{TenBarge2014}). Since $B_{\infty y}$ decreases for increasing $b_g$, it becomes more comparable to the noise level for very high $b_g$. Then,  for large $b_g$ the signal-to-noise ratio of all the fields can be very low, as can be seen in the extreme case $b_g=50$ in Fig.~\ref{fig:pth}(e) and Fig.~\ref{fig:bz_contours}(e) (even with the extended temporal average). Therefore, although in principle there is no limitation due to numerical constraints on the electron gyromotion for even higher guide fields (in our setup, $\rho_e$ and $\Omega_{ce}$ are constant for different guide fields), the PIC results in this regime are not reliable due to this unfortunate fact.

\section{Transverse distance}\label{sec:app_transverse_distance}

\begin{figure}[h!]
	\centering
	\includegraphics[width=\linewidth]{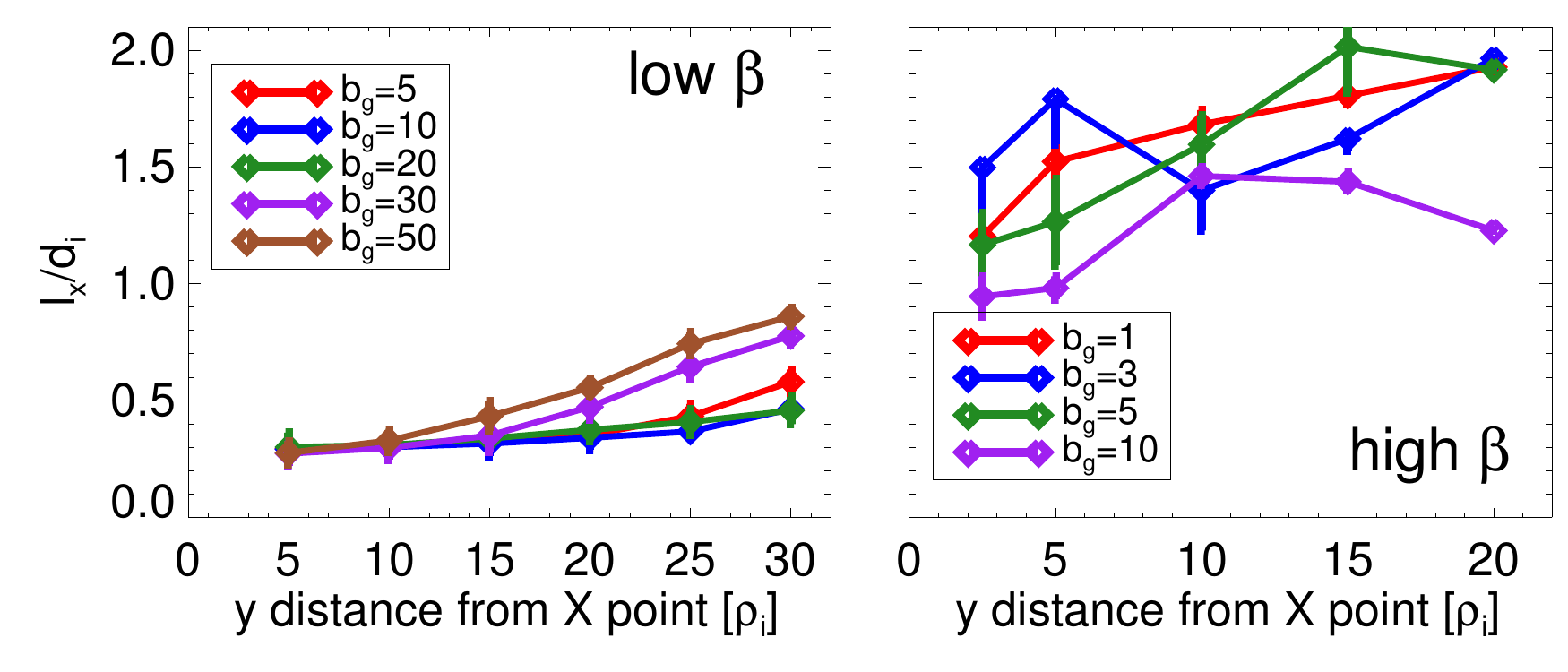}
	\caption{Estimation of the transverse distance $l_x/d_i$ for both low (left panel) and high  (right panel) beta case and different guide fields. These values were taken from plots of the thermal pressure $\delta P_{th}$ after the reconnection peak time $t=50\tau_A$ shown in Fig.~\ref{fig:pth} (for the low beta case). The specific method is explained in the text. \label{fig:transversal_distance_scaling}}
\end{figure}

Here, we proved the statement written in Sec.~\ref{sec:symmetry}: the transverse distance of the thermal and magnetic fluctuations scale as $l_x/d_i \sim \sqrt{\beta_i}$. This implies: (1) it is constant for different guide fields (requirement for the inverse scaling with $b_g$) and (2) it is higher for the high than the low beta set of parameters. We estimated the value of $l_x/d_i$ shown in Fig.~\ref{fig:transversal_distance_scaling} by using the plots of $\delta P_{th}$ in Fig.~\ref{fig:pth}. First, we detected the main X point by locating the minimum of the vector potential $A_z$ along the center of the CS. Then, we chose several distances to the left of that point in the $y$ direction along the center of the CS ($l_y$, indicated in the $x$ axis of Fig.~\ref{fig:transversal_distance_scaling}). From each point we measured the transverse distance across the $x$ direction, $l_x$, from the center until the point in which the current density $J_z$ drops to $20\%$ and $10\%$ of its initial peak value, as a simple way to detect the boundaries of the CS in these regions (averaging in both positive and negative $x$ directions). These values correlate well enough with the approximate boundaries of the regions with significant values of $\delta P_{th}$ (as well as $\delta B_z$) above numerical noise, besides of being independent of scaling arguments. The small error bars are the differences between the $10\%$ and $20\%$ of the initial value of $J_z$.
First of all, in Fig.~\ref{fig:transversal_distance_scaling}(left) we can see that for the low beta case, the transverse distance is more or less constant for different guide fields at least for (longitudinal) distances $l_y<15\rho_i$ away from the main X point,  confirming the assumption stated before. As can be expected, the transverse distance $l_x$ increases away from the X point, consequence of the more open separatrices  but also because the distinction between them and the boundary of the main magnetic island is more diffuse. In this region, there are more deviations from the previous constant trend for $l_y<15\rho_i$ among different guide fields, but in any case, they are not significant, and a value $l_x\sim 0.3\,d_i$, corresponding to 3 times the order of magnitude estimate  $l_x/d_i \sim \sqrt{\beta_i}=0.1$, can be considered representative not too far away from the X point. Note that the deviations from more or less constant values are specially significant for the cases of higher guide field $b_g=30$ or $b_g=50$, since the structures of the separatrices away from the X point and close to the O points are different from the low guide field cases (see Fig.~\ref{fig:pth}). On the other hand, the estimations for the transverse distance for the high beta case in Fig.~\ref{fig:transversal_distance_scaling}(right) show larger variations among guide fields, since in many cases the CS develop small secondary islands to the left and very close of the main X point and so the measurement is biased. In addition, for this high beta case, the enhanced level of numerical noise makes the detection of the transverse distance $l_x$ more inaccurate.  For this case, it is only possible to conclude that the transverse distance is in between a certain range with a spread of $\Delta l_x/d_i\sim 0.5$. Nevertheless, close enough to the X point ($l_y<5d_i$), $l_x=1.3-1.5\,d_i$ can still be considered reliable enough for a comparison (excluding the extreme case $b_g=10$). This value turns out to be practically, in order of magnitude, the estimate $l_x/d_i \sim \sqrt{\beta_i}=1.0$.

Therefore, from the results shown in Fig.~\ref{fig:transversal_distance_scaling}, we can confirm that   $l_x/d_i $ is (1) more or less constant for different guide fields, especially in the low beta case, and (2) its order of magnitude is between 1-3 times $\sim \sqrt{\beta_i}$, which is good enough to ensure the validity of the inverse scaling with $b_g$ of  $\delta P_{th}$ and  $\delta B_{z}$.  Note that if we had chosen to keep $B_{\infty y}$ constant and increase $B_g$ to have a higher guide field effect,  the total $\beta_i$ would not be constant and so the inverse scaling, implying that a direct comparison between the different PIC guide fields and GK runs would not be possible.

\section{Computational performance of the GK/PIC codes}\label{sec:app_times}

It is worthwhile to mention the speed-up of the GK simulations compared to the PIC runs,  the main motivation of using the first numerical method over the second one for magnetic reconnection studies. Because of the choice of keeping the total plasma $\beta$ constant for different PIC guide fields to properly compare with the GK results (see  Appendix~\ref{sec:app_normalization}), the Alfv\'en time  in units of $\omega_{pe}^{-1}$ is (practically) linearly proportional to $b_g$ according to
		\begin{equation}\label{tau_omega}
		\tau_A\omega_{pe}=\frac{\sqrt{T_i/T_e}m_i/m_e}{v_{th, e}/c}\beta_i\frac{L}{\rho_i}\sqrt{1+b_g^2}.
		\end{equation}
In the PIC runs,  because the time step has to be proportional to $\omega_{pe}^{-1}$ for stability reasons, the latter expression will also be proportional to the number of time steps used to reach a given time measured in $\tau_A$,  and thus to the computational effort. Then,  PIC high guide field runs are computationally more expensive than the runs in the low guide field regime (as well as the high beta cases compared to the low beta ones). In fact, for the low beta case, the ACRONYM PIC code used $3.38\cdot10^4$ CPU core-hours to run the cases $b_g=5$ up to $\tau_A=70$ (the last time shown in Fig.~\ref{fig:recrates}),  while
 $3.33\cdot10^5$ CPU core-hours for the case $b_g=50$. A significant fraction of this computational effort is spent in running time diagnostics for higher order momenta of the distribution function, which are used in the results to be shown in a follow-up paper. On the other hand,  the single GENE GK code simulation with which those runs were compared used only $350$ CPU core-hours,  representing a speed-up by a factor of $10^3$ when comparing with the largest PIC guide field considered $b_g=50$. These huge computational savings are an additional justification for the  importance of a proper comparison study of GK with PIC simulations of guide field reconnection,  one of the purposes of the present paper.

\end{document}